%% file: paper.tex
\documentclass[nofootinbib,twocolumn,prd]{revtex4}
\usepackage{hyperref}
\usepackage{amsmath}
\usepackage{graphicx}
\usepackage{color}

\newcommand\richardBugFix{\textbf{\color{red}Richard will bugfix}}
\newcommand\editremark[1]{ {\color{red} #1}}

\newcommand\optional[1]{}
\newcommand\ForInternalReference[1]{}

\newcommand\unit[1]{\, {\rm #1}}

\newcommand\qmstateproduct[2]{\left<#1|#2\right>}

\newcommand\myvector[1]{{\mathbf{#1}}}
\newcommand\chip{{\boldsymbol{\chi}_+}}
\newcommand\chim{{\boldsymbol{\chi}_-}}

\newcommand\wStarNonspinning{1.095}

\newcommand{\ROS}[1]{}

\def\bbh#1{binary black hole#1 (BBH#1)\gdef\bbh{BBH}}
\def\bh#1{black hole#1 (BH#1)\gdef\bh{BH}}

\begin{document}
\title{Intrinsic selection biases of ground-based gravitational wave searches for high-mass BH-BH mergers} 
\author{R.\ O'Shaughnessy}
 \affiliation{
 Center for Gravitational Wave Physics, Penn State University,
 University Park, PA 16802, USA}
\altaffiliation{Current address:
Center for Gravitation and Cosmology, University of Wisconsin-Milwaukee,
Milwaukee, WI 53211, USA}
\email{oshaughn@gravity.phys.uwm.edu}
\author{B. Vaishnav}
\affiliation{
Center for Gravitational Wave Astronomy,  
The University of Texas at Brownsville,
80 Fort Brown, Brownsville, Texas 78520 USA
}
\author{ J. Healy}
\author{D. Shoemaker}
\affiliation{Center for Relativistic Astrophysics,
Georgia Tech, Atlanta, GA 30332, USA}

\begin{abstract}
The next generation of ground-based gravitational wave detectors may detect a few mergers  of comparable-mass $M\simeq 100-1000 M_\odot$
(``intermediate-mass'', or IMBH) spinning black holes.    Black hole spin is known to have a significant impact on the orbit,
merger signal, and post-merger ringdown of any  binary with non-negligible spin.  In particular, the detection volume for spinning binaries
depends significantly on the component black hole spins.
We provide a fit to the single-detector and isotropic-network detection volume versus (total) mass and arbitrary spin for equal-mass binaries.  Our analysis
assumes matched filtering to all significant available waveform power (up to $l=6$ available for fitting, but only $l\le 4$ significant) estimated by an array of 64
numerical simulations with component spins as large as $S_{1,2}/M^2\le 0.8$.  We provide a spin-dependent estimate of our
uncertainty, up to $S_{1,2}/M^2\le 1$.  For the initial (advanced) LIGO detector, our fits are reliable for 
$M\in[100,500]M_\odot$ ($M\in[100,1600]M_\odot$).  In the online version of this article, we also provide
fits assuming incomplete information, such as the neglect of higher-order harmonics.  We briefly
discuss how a strong selection bias towards aligned spins influences the interpretation of future gravitational wave
detections of IMBH-IMBH mergers.
\end{abstract}
\maketitle

\section{Introduction}
Ground-based gravitational wave detectors like LIGO and Virgo are presently taking data at and beyond design
sensitivity \cite{gw-detectors-VIRGO-original,gw-detectors-LIGO-original}.  Over the next  several
years as upgrades are performed, these detectors' sensitivity will increase substantially
\cite{LIGO-aLIGODesign-Sensitivity}.  Advanced detectors are very likely to see many  few-stellar-mass black hole binaries
formed through isolated \cite{PSellipticals,popsyn-LowMetallicityImpact-Chris2010} and dynamical
\cite{clusters-2005,2008ApJ...676.1162S,2010MNRAS.402..371B} processes.  Additionally, advanced detectors could see the merger signature of
two \emph{intermediate-mass black holes} (each $M\in[100, 10^3]$), binaries which might be formed in dense globular
clusters \cite{imbhlisa-2006}.  
Unless astrophysical processes strongly suppress black hole spin, that spin will have a substantial effect on all the
components of the signal to which these ground-based detectors are sensitive:  the
late-time inspiral, merger signal, and (through the final BH spin) ringdown. 
Though analytic approximations exist to describe the early-time (inspiral) and late-time (ringdown) behavior of
a spinning BH-BH binary,  the merger signal must be obtained numerically, in principle for all possible mass ($m_1,m_2$)
and spin ($\myvector{S}_1,\myvector{S}_2$) combinations.  Of the types of black holes likely to be detected in the near future, intermediate-mass
black holes have masses and spins such that their \emph{entire} (short) detectable waveform is dominated by the merger
signal.
As a result, very few candidate intermediate-mass merger waveforms are presently available, particularly for generic spins~\cite{2010PhRvD..81h4023L,2009PhRvD..79h4010C,2009PhRvD..80l4010S,2010CQGra..27h4034C}.
Conversely, the performance of real gravitational wave search pipelines is difficult to assess without extensive Monte
Carlo simulations.  All searches suffer from  highly nongaussian noise and time-variable detector performance;  matched
filter searches adopt a range of approximate waveforms, coincidence, and nongaussian-noise rejection strategies.
However, given the computational burden of each  NR simulation and the limited selection currently available, extensive
Monte Carlo studies are not presently practical.
Nonetheless, all numerical simulations to date suggest that gravitational merger waveforms are \emph{surprisingly simple}. 
For example, for aligned spin both the final spin \cite{gr-nr-io-review-Rezzolla2008,2008PhRvD..77f4010M,gr-nr-io-fitting-BR2009,2010PhRvD..81h4054K} and even merger waveforms
\cite{gwastro-Ajith-AlignedSpinWaveforms,gwastro-nr-AlignedSpinVolumeWeight,2010PhRvD..81h4041P} have been accurately fit, for all possible
component masses $(m_1,m_2)$ and spin magnitudes  $|\myvector{S}_1|,|\myvector{S}_2|$ for moderate spin magnitude
($<0.9$) and mass ratio ($<1/10$), for the dominant mode of radiation.

Given the simplicity of numerical merger waveforms and the few  available simulations, in this paper we outline a simple
method to \emph{analytically} estimate the performance of present and future gravitational wave searches, extrapolating
from a small array of existing numerical simulations.
Our method relies only on the raw numerical simulation output and detector response functions; we neither  model 
the waveform itself nor limit to a few ``dominant harmonics'' associated with the early- or late-time binary orientation.\footnote{In other words, we neither fit to $h$ of t or f nor do we pick a
  single ``dominant mode'' like $h_{22}$, which relies on a preferred frame.
}
In Sec. \ref{sec:Principles} we describe  (i)  how we estimate the detection volume of present and future
gravitational wave searches for comparable mass binaries with known waveforms and (ii)  how we extrapolate between
them using fits.
In Sec. \ref{sec:NR} and Table \ref{tab:SimulationSet} we describe the set of NR simulations used.
In Sec. \ref{sec:Results:AlignedSpin}, we compare our results on aligned spin to previously published estimates.
Additionally, we use our aligned spin results to emphasize how sensitive our predictions are to small systematic issues
such as  wave extraction radius and (to a lesser extent) numerical resolution.
Then, in Sec. \ref{sec:Results:Generic} we provide the coefficient functions needed for arbitrary spins with total and
incomplete ($l_{max}=2,3,4,\ldots$) waveform catalogs, along with our
best estimates for parameter-dependent uncertainty in the detection volume.
Finally, in Sec. \ref{sec:Astro} we discuss how much (or little) spin influences searches for high-mass IMBH-IMBH mergers.

\section{GW Searches for high-mass mergers: Selection biases }
\label{sec:Principles}

The sensitivity of a network of gravitational wave detectors to a single class of randomly oriented source is often
characterized in three ways: (A) via the maximum amplitude an
optimally oriented but otherwise identical  binary produces; (B) using the
angle-averaged signal power $\bar{\rho}_*$ incident on a single detector \cite{1998PhRvD..57.4535F} from sources at a
fixed distance;  (C) via the \emph{expected detection rate} for that
class of source \cite{PSellipticals}, adopting a distribution $p(\lambda)$ of sources
described by parameters $\lambda$ ($=$ component masses $m_1,m_2$; spins $\myvector{S}_1,\myvector{S}_2$;  emission direction $\myvector{\hat{n}}$ in the
frame of the binary;
sky location and polarization angle $\hat{N},\psi$; and distance $r$).
For example, for a single  interferometer with Gaussian noise, the signal-to-noise ratio $\rho_*(\lambda)$ can be expressed as
\begin{eqnarray}
h_{det}(t) &=& F_+(\hat{N},-\psi) h_+(\myvector{\hat{n}},t) + F_\times(\hat{N},-\psi) h_\times(\myvector{\hat{n}},t) \nonumber \\
\qmstateproduct{a}{b} &\equiv& 2\int_{-\infty}^\infty df \frac{\tilde{a}(f)^* \tilde{b}(f)}{S_h(f)} \\
\rho_*^2 &=& \qmstateproduct{h_{det}}{h_{det}}
\end{eqnarray}
where $F_{+,\times}$ are standard single-detector beampattern functions and where we use a subscript $*$ when referring
to a single detector; compare to Appendix \ref{ap:AlternateNetworks}.
For the first method, a \emph{peak} signal to noise $\rho_{*,max}$, is particularly useful for \emph{aligned or nonspinning binaries}
dominated by $l=|m|=2$ emission,
where the relative change in $\rho$ versus $\hat{N},\psi,\myvector{\hat{n}}$ is known analytically; see, e.g., Eq. (2) in
\cite{PSellipticals}.  
In the second method, the source- and sky-location-averaged signal power incident on a single detector leads naturally to
orientation-averaged power over a \emph{complex} wave amplitude
\begin{eqnarray}
\label{eq:def:rhobar:star}
\bar{\rho}_*^2 &=& \left< \qmstateproduct{ h_{det}}{h_{det}} \right>_{\psi,\hat{n},\hat{N}} 
  \equiv \int \frac{d\Omega_N d\Omega_n}{(4\pi)^2} \frac{d\psi}{\pi} 
 \qmstateproduct{ h_{det}}{h_{det}}
 \\
&=& \nonumber 
\int \frac{d\Omega_n}{4\pi} 
 \int \frac{ d\Omega_N d\psi}{4\pi^2}  \left[  F_+^2 \qmstateproduct{h_+}{h_+} +0+
   F_\times^2\qmstateproduct{h_\times}{h_\times} 
  \right] \\
&=& \frac{1}{5} \int \frac{d\Omega_n}{4\pi} [\qmstateproduct{h_+}{h_+} + \qmstateproduct{h_\times}{h_\times}] \nonumber
\\
 &=& \frac{1}{5} \bar{\rho}^2 
\end{eqnarray}
where $\bar{\rho}^2$ is a technically convenient lower-dimensional average with clear physical meaning -- the average
over all \emph{source} orientations of the network signal-to-noise  recovered by either (i) a \emph{pair}
of identical detectors, oriented at $45^o$ to each other and with the source directly overhead, or equivalently (ii) a
network of detectors with equal sensitivity to both polarizations in all directions [cf. Appendix \ref{ap:AlternateNetworks}]:
\begin{eqnarray}
\label{eq:def:rhobar}
\bar{\rho}^2 &\equiv& \int\frac{d\Omega_n}{4\pi} \qmstateproduct{h_++i h_\times}{h_++i h_\times}_{\text{fixed }\hat{n}} \;,
\end{eqnarray}
Thus, the orientation-averaged  power is technically convenient since, if the emitted waveform is expressed as an expansion of the asymptotic complex waveform $h_++ih_\times$ or 
curvature scalar $\Psi_4$ 
into spin-weighted spheroidal harmonics
\begin{eqnarray}
h_++i h_\times &=& \sum_{lm} h_{lm}(t) \,  {}_{-2}Y_{lm} \,,
\end{eqnarray}
then the angle-averaged signal power becomes a sum over inner products of the harmonic amplitude functions $h_{lm}$
or (with a different inner product) $\Psi_{4,lm}$ given by
\begin{eqnarray}
\bar{\rho}^2&=&  \sum_{lm} \frac{\qmstateproduct{h_{lm}}{h_{lm}}}{4\pi} %
 \\ 
 & =&  \sum_{lm} \frac{1}{2\pi r^2}  \int_0^\infty \frac{df}{(2\pi f)^4S_h} 
\left[ 
  |\tilde{\Psi}_{4,lm}(f)|^2   +   |\tilde{\Psi}_{4,lm}(-f)|^2  
  \right] \nonumber  \\
\label{eq:rho:bar:psi4sum}
&\equiv& \sum_{lm} (\Psi_{4,lm}|\Psi_{4,lm})\,;
\end{eqnarray}
see, for example, Eqs. (7,8) in \citet{gwastro-nr-AlignedSpinVolumeWeight}.
For clarity we have described both methods (A) and (B)  as a characteristic SNR $\rho$ for sources at a fixed distance, adopting a
detection-strategy-neutral characterization.  If one adopts a fiducial signal to noise ratio $\rho_c$, such as a cutoff
for single-detector SNR, then these amplitudes convert to physical distances; for instance, (B) implies an
angle-averaged reach $\bar{D}_*$ defined by the solution to $\bar{\rho}_*(\bar{\lambda},D)=\rho_c$, or equivalently by
\begin{eqnarray}
\bar{\rho}_*(\bar{\lambda})&=& \rho_c \frac{\bar{D}_*(\bar{\lambda})}{r}
\end{eqnarray}
where $\bar{\lambda}=(m_1,m_2,\myvector{S}_1,\myvector{S}_2$ and suitable orbital phases)  are the physical (``intrinsic'') parameters of the
binary (i.e., all parameters except $\psi,N,n$ and distance). 
Though well-defined, both methods only approximate the astrophysically-relevant sensitivity of gravitational wave detectors to spinning
systems.  For example,  both the distribution and even optimal emission direction  ($\psi,\myvector{\hat{n}}$) depend strongly on the direction of the total angular momentum $J$ in band; and,
therefore, on the masses and spins involved.    
While the \emph{peak} amplitude could  easily be tabulated and fit
following the procedure described below, for astrophysical purposes (A) and (B) alone lose information about the
beampattern shape function needed to construct (C), the \emph{astrophysically}-relevant measure of sensitivity.  
To encapsulate all needed information about the beampattern shape, we introduce a beampattern function $w_*$ for the
ratio of single-detector SNR $\rho$ to orientation-averaged single-detector SNR $\bar{\rho}_*$: %
\begin{eqnarray}
\label{eq:def:w}
w_* &\equiv&  \rho_*(\lambda)/\bar{\rho}_*(\bar{\lambda})  \\ %
\rho_* &=& \frac{\rho_c \bar{D}_*(\bar{\lambda})w_*(\hat{N},\myvector{\hat{n}},\psi|\bar{\lambda})}{r} \,.
\end{eqnarray}
By construction, the orientation average of $w^2_*$ is always exactly unity ($\left<w^2_*\right>=1$).
In terms of this beampattern function and the previously defined angle-averaged range $\bar{D}$, the detection rate for
sources in the nearby universe can be expressed as a sum over the rate per unit physical volume $dV=r^2 dr d\Omega_N$ and per volume in binary
parameters $d\lambda = d\Omega_{source}d\bar{\lambda}$:
\begin{eqnarray}
R_D &=& \int_{\rho> \rho_c} \frac{dN}{dt dV d\lambda} dV d\lambda \nonumber \\
& =& \int \frac{dN}{dt dV} p(\bar{\lambda}) d\bar{\lambda} \int \frac{d\Omega_{source}}{\Omega_{source}}\int_{\rho_*>\rho_c} dV \,.
\nonumber 
\end{eqnarray}
The last factors represent the detection volume averaged over source orientations.  In the nearby universe, this
orientation-averaged detection volume can be evaluated, reducing the detection rate to
\begin{eqnarray}
\label{eq:DetectionRate}
&=&  \frac{dN}{dt dV} \int p(\lambda) d\lambda \; \frac{4\pi}{3}
(\bar{D}_* \bar{w}_*)^3 \,, 
\end{eqnarray}
which shows that the final detection rate can be explained as a product of: (i) total
event rate per unit volume $\frac{dN}{dt dV}$; (ii) distribution of events in parameters; and a (iii)
source-frame-averaged volume that characterizes the typical reach to sources with parameters $\lambda$, consisting of
(iv) an \emph{orientation-averaged range} $\bar{D}_*$, proportional to the band-limited SNR
$\bar{\rho}_*$ for a source at a fixed distance, and (v) a \emph{beaming correction factor}
$\bar{w}_*$, that depends on how the binary's polarized, beamed emission interacts with our network's polarization-dependent beampattern.
In other words, the angle-averaged reach $\bar{D}_*$ \emph{almost} describes the astrophysically relevant reach, modulo a
weak correction factor 
\begin{eqnarray}
\label{eq:def:wbar:star}
\bar{w}_*(\bar{\lambda})&\equiv&  \frac{\int \frac{d\Omega_n d\Omega_N d\psi}{4 \pi (4\pi)( \pi)} 
\left[ w_*(\myvector{\hat{n}},\psi,\hat{N}|\bar{\lambda})^3\right ]^{1/3}
}{\left<w^2_*\right>^{2/3}}
\end{eqnarray}
which, like $\bar{\rho}$, can be tabulated and fit versus all intrinsic parameters $\bar{\lambda}$.
As we will see below, the beampattern correction function average  $\bar{w}_*$ is necessarily  almost always
unity, with $\bar{w}_*-1$ largest for aligned binaries and exactly zero for isotropic emission.  In fact, given the uncertainties expected in our fit to
$\bar{\rho}(\lambda)$, an excellent first approximation  to  $\bar{w}_*$ is  unity.

\subsection*{Fit versus spin}
Owing to the relative simplicity of numerical merger waveforms, scalar functionals of the waveforms like the final mass
$M_f$ and spin $J_f^2$ can be fit across
the entire space of intrinsic parameters $\bar{\lambda}$ \cite{2008PhRvD..78b4017B}.  Fits for the remnant BH's  recoil kick and spin have been extensively
explored in the literature
\cite{2008PhRvD..78b4017B,gr-nr-io-fitting-Boyle2007,2008PhRvD..78d4002R,2008ApJ...674L..29R,gr-nr-io-fitting-GoddardGeneric2008,gr-nr-io-review-Rezzolla2008,gr-nr-io-fitting-BR2009}.
Experience from  fitting the final spins and signal-to-noise-ratio of \emph{aligned} spinning binaries
[Sec. \ref{sec:Results:AlignedSpin}] suggests an accurate fit to $\bar{\rho}(\myvector{S}_1,\myvector{S}_2)$ requires at least \emph{cubic} order in the
components of $S_{1,2}$ along the early-time orbital angular momentum direction $\hat{L}$  (henceforth denoted
$\myvector{\hat{z}}$).\footnote{To avoid ambiguity,  I adopt the initial orbital angular momentum $\hat{L}$ at $r=6.2$ as a reference
  direction; see Section
  \ref{sec:NR}.     \ROS{Request for comment}}
Though a generic symmetry-preserving expansion contains many components (roughly $6^3/4$ in a cubic-order expansion of a
six-dimensional space with
two $Z_2$ symmetries, parity and black hole exchange\footnote{Following \cite{2008PhRvD..78b4017B} a generic scalar expansion
  would have 1,2,11, and 23 parameters at zeroth, first, second, and cubic orders, respectively.  Not including the
  trivial zeroth order term, our proposed expansion has
  4 fewer parameters at quadratic order and 12 fewer parameters at cubic order.}), the physics of a precessing merging binary strongly suppresses
most terms.  A truly generic symmetry preserving expansion of a scalar function allows  scalar functions of $\myvector{S}_1,\myvector{S}_2$
to have a \emph{preferred spin direction} perpendicular to the total angular momentum, along the axis connecting the two holes at our
simulations' starting time.
Precession rapidly evolves  in-plane spin components away  from this preferred axis.     
We anticipate and test simulations confirm minimal dependence of the amplitude $\bar{\rho}$ on the relative orientation of spins to
this preferred axis. 

\ForInternalReference{
For simplicity, I do not attempt
to deduce the true direction of the initial total angular momentum $\hat{J}$ and use it as a reference direction.  Nor
do I use the final BH spin direction for reference (presumably similar).
}

\ForInternalReference{

\begin{figure}
\includegraphics{fig-mma-ConsistencyTest-PerpendicularDirectionIrrelevant}
\caption{\textbf{Consistency test: James' simulations}: The triangle test.  
Top panel: $\rho(M)$ Solid lines: James simulations ($l=2$ only) shown for
  $r=60$.  These simulations have $a_2 = 0.6\hat{z}$ and $a_1$ changing along a triangle at $\theta=\pi/4$.  They thus
have identical $z$ and $|P$ components of $\chi_\pm$ (i.e., $\chi_{\pm z}=0.3(\pm 1 +  1/\sqrt{2})$, $|P\chi_\pm| = |P
a_1/2| = 0.3/\sqrt{2}\simeq 0.21$, though the direction of the
  perpendicular components varies.  Filled region: predicted amplitudes for $\rho(M)$, including error interval, using
  the standard fit and error estimate; compare with Figure \ref{fig:Aligned:Test:Range}
Bottom panel: the l=m=2 mode for $|\psi_4(t)$ for the 3 simulations.  Small differences do exist.
}
\end{figure}

}

For this reason, rather than use all spin components, as in \citet{2008PhRvD..78b4017B}, we describe our expansion in
terms of three quantities: the preferred out of plane direction $\myvector{\hat{z}}\propto J$;  the in-plane projection operator $P$ perpendicular to
$\myvector{\hat{z}}$; and a pair  even- and odd- exchange-symmetric reduced spins:
\begin{eqnarray}
\boldsymbol{\chi}_\pm  &=& (m_1 a_1 \pm m_2 a_2)/M  \qquad [\boldsymbol{\Sigma} \propto  - \chim] 
\end{eqnarray}
where $a_k=S_k/m_k^2$.   An exchange-symmetric expansion of equal-mass binaries must have even powers of $\chim$.  [For
unequal masses, terms proportional to $\chim$ are possible  when prefixed by asymmetric mass terms.  A generic expansion
will introduce at linear spin order $(\delta m/M)(\chim\cdot\myvector{\hat{z}})$, at quadratic spin order  $(\delta
m/M) P \chim\cdot P\chip$,  et
cetera.  The exchange-symmetric coefficients provided below can depend on only even powers of $(\delta m/M)$.]
Working to cubic order, we anticipate $\bar{\rho}$ has the form
{\small \begin{eqnarray}
\label{eq:expansion:barrho}
\bar{\rho}(m_1,m_2, \myvector{S}_1,\myvector{S}_2)&=& \bar{\rho}_o(m_1,m_2)[1    \\
 &+& {\cal X}_1 (\chip\cdot z) + {\cal X}_2  (\chip\cdot \myvector{\hat{z}})^2 + {\cal X}_3  (\chip\cdot \myvector{\hat{z}})^2 
\nonumber \\
 &+&  {\cal X}_{02}  (P \chip)^2  +{\cal A}_{20}(\chim\cdot z)^2 + {\cal A}_{02}(P \chim)^2 
 \nonumber \\
 &+ & {\cal B}_{1200} (\chip \cdot z) (P \chip)^2 
\nonumber \\ &+&  {\cal B}_{1020} (\chip \cdot z) (\chim \cdot z)^2
 \nonumber \\
 &+&  {\cal B}_{1002}(\chip \cdot z) (P\chim)^2 + O(\chi^4)
\ldots]  \nonumber \\
\label{eq:expansion:barw}
\bar{w}_*(m_1,m_2,\myvector{S}_1,\myvector{S}_2)&=&  \bar{w}_{o,*}(m_1,m_2)[1 \\
&+& {\cal Z}_1\chip\cdot \myvector{\hat{z}}+ {\cal Z}_2(\chip\cdot \myvector{\hat{z}})^2  \ldots   \nonumber \\
&+& {\cal C}_{20}(\chim\cdot \myvector{\hat{z}})^2 + {\cal C}_{02}(P\chim)^2 + \ldots \nonumber  \\ 
&+& {\cal D}_{1200}(\chip\cdot z)(P\chim)^2 +\ldots \nonumber \\
&=& \wStarNonspinning[1+{\cal Z}_1\chip\cdot \myvector{\hat{z}} + \ldots] \nonumber %
\end{eqnarray} }
\noindent where $\bar{\rho}_o(m_1,m_2)$ is provided by a comparable calculation for nonspinning binaries and where the
value for $\bar{w}_{o*}$ for $\myvector{S}_1=\myvector{S}_2=0$ shown can be estimated using only the dominant $l=|m|=2$ aligned emission
beampattern.\footnote{To an excellent approximation the beampattern factor $w$ agrees with the average associated with
  $l=|m|=2$ emission and can be calculated by angle-averaging  the analytically-known form of $w$ for that case; see
  \citet{PSellipticals}, noting $w$ there differs by a constant factor from our definition.
}
In most of the text we will refer explicitly to the coefficient functions described above.  However,  when
referring to this expansion in its entirety, we adopt the shorthand abstract notation  $y_\alpha$ for its
coefficient functions (of $M$)
and  $\psi_\alpha$ for basis functions (of $\chi_{\pm}$):
\begin{eqnarray}
\bar{\rho}/\bar{\rho}_o = \sum_\alpha \psi_\alpha y_\alpha\,.
\end{eqnarray}

For equal-mass binaries $m_1=m_2=M/2$, we determine the coefficient functions ${\cal X}_{1,2,3}(M),\ldots$ suitable to
a specific gravitational wave detector noise power spectrum $S_h$ as follows.\footnote{We  adopt an the initial LIGO sensitivity from
  \cite{LIGO-Inspiral-S5-Ranges} and an  advanced detector noise power spectrum from 
\cite{LIGO-aLIGODesign-Sensitivity}.
}  We pick a total mass $M$.  For all numerical simulation $k$ in Table \ref{tab:SimulationSet}, corresponding to spin combinations
$S_{1,k},S_{2,k}$, we extract time domain 
spin-weighted harmonics of the Weyl scalar $\Psi_{4,lm,k}(t)$ for $l\lesssim 6$.   We construct each $\bar{\rho}_k$
using Eq. (\ref{eq:rho:bar:psi4sum}).  Likewise, we construct $\bar{w}$ by (i) reconstructing $\tilde{h}$ along a large number of
randomly chosen orientations ($\hat{N},\psi,\myvector{\hat{n}}$), using the multipole coefficients $\tilde{\Psi}_{4,lm}$; (ii)
  calculating $w_*^3$; then (iii) averaging over all the random samples.  
Finally, excepting only a few simulations chosen as blind tests of our fit [Sec. \ref{sec:sub:BlindTest}], we perform a simple least-squares fit\footnote{Though we prefer a global fit to all
  parameters, the individual coefficient functions themselves can be (nearly) isolated using suitable-symmetry  one-parameter
  subfamilies of simulations.  For example, the coefficient ${\cal A}_{20}$ can be determined from simulations with antialigned spins
  $\myvector{S}_1=-\myvector{S}_2=|S| \myvector{\hat{z}}$; the coefficient ${\cal B}_{1020}$ from simulations with aligned but unequal spins, given
  knowledge of equal-spin coefficients; the
  coefficient ${\cal A}_{02}$ can be determined  from the ``B-series'' \cite{Herrmann:2007ex}, where both spins are antialigned ($\chip=0$)
  and tilted at a range of angles $\theta$.  For brevity, we only discuss a global fit, rather than fits to individual simulation subfamilies.
} for the coefficients $y_\alpha$ in
Eq. (\ref{eq:expansion:barrho}), adopting uniform uncertainties in  $\rho_k$ for all $k$.
Since $\bar{w}$ varies little from unity and
  since our estimate is subject to significant Poisson sampling error, we retain only leading-order dependence with spin
  (${\cal Z}_1$).

\noindent \emph{Fit errors}: Numerical, systematic,  and truncation  errors limit our ability to determine these coefficients
precisely.  High-order or symmetry-suppressed coefficients like ${\cal X}_3$ and ${\cal A}_{20}$ are particularly sensitive to small numerical
errors.  Furthermore, the recovered coefficients $y_\alpha(M)$ have highly correlated uncertainties
$\Sigma_{\alpha\beta}(M)$, where $\Sigma_{\alpha \beta}$ is the least-squares estimate of the parameter covariance
matrix.  
Finally, phenomenologically speaking the interesting uncertainty is how much our  \emph{fit} $\hat{F}$ to our data, say
for $\bar{\rho}$,
\[
F_k = \bar{\rho}_k/\bar{\rho}_o \, ,
\]
differs from truth.  Though highly dependent on the parameter distribution of simulations used in the fit, one simple
estimate for overall error is the residual rms error between the data and our least-squares fit:
\[
\delta F^2 = \sum_k (\hat{F}(\lambda_k)-F_k)^2/N\,.
\]
 A far more stable and spin-magnitude-dependent representation of fit uncertainty  is the expected $L^2$ error ${\cal F}(a)$ in the fit for spins with
magnitude $\myvector{S}_1/m_1^2,\myvector{S}_2/m_2^2\le a$:
\begin{eqnarray}
\label{eq:err:J}
{\cal F}^2(a)&=& \left<(\bar{\rho} - \left<\bar{\rho}\right>)^2\right>_{fit}/\bar{\rho}_o^2 \nonumber \\
 &\equiv & \sum_{ab} \Sigma_{ab}\left<\psi_a\psi_b\right>_{spins} a^{s_a+s_b}
\end{eqnarray}
where we use integer exponents $s_a$ to describe the order of the basis functions ($\psi_a(x \myvector{S}_1,x\myvector{S}_2)=x^{s_a}
\psi_a(\myvector{S}_1,\myvector{S}_2)$); where $\Sigma_{\alpha,\beta}$ is the least-squares covariance matrix of fit parameters; and 
 where the coefficients are easily-tabulated moments of the coefficient matrix over all spins, provided in Table
 \ref{tab:ErrorDiagnostics:MomentMatrix}:
\begin{eqnarray}
\label{eq:err:CoefMatrix}
\hspace{-0.6cm}\left<\psi_a\psi_b\right>_{all} &=& \int_{|a_{1,2}|<1} \frac{d^3a_1 d^3a_2}{(4\pi/3)^3} \psi_a(a_1,a_2) \psi_b(a_1,a_2) \,. 
\end{eqnarray}
Roughly speaking, ${\cal F}(a)$ estimates the  relative error in $\bar{\rho}$ when applying our fit to a generic pair of
spins with typical magnitude $a$.

\begin{figure}
\includegraphics{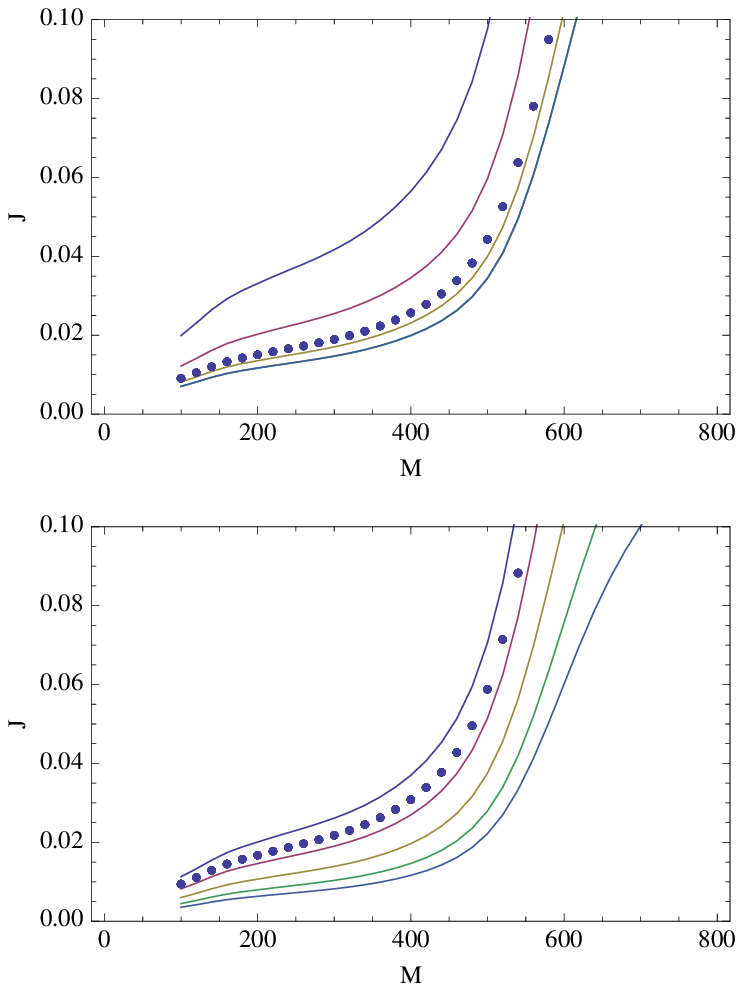}
\caption{\label{fig:ErrorDiagnostics-GlobalVsM-Cubic} Top panel: Expected $1\sigma$ error ${\cal F}$ in an unconstrained
  fit to 
  $F=\bar{\rho}/\bar{\rho}_o$, for spin magnitudes    $|a_1|,|a_2|$ limited to below unity (blue; largest); below $0.8$
  (red); and below  $0.6, 0.4$ and $ 0.2$ (similar
  curves near bottom).  Also shown (points) are the least-squares errors $\delta F$ between our fit $\hat{F}$ and the
  simulated NR data.
Bottom panel: As above, but using a model where only ${\cal X}_{1,2,3}$ and ${\cal X}_{02}$ are nonzero.
}
\end{figure}

\noindent \emph{Astrophysical tolerance}:  Astrophysically speaking, the orientation-averaged range
relates the number of sources observed (or their absence) to the implied source event rate (or
upper bound).   Even adopting the most optimistic assumptions,  no more than $O(10)$ intermediate-mass mergers are
expected to be seen by advanced ground-based detectors \cite{imbhlisa-2006}, at best allowing the source event
rate be determined to  $O(30\%)$ at $1\sigma$ confidence.   To achieve this level of accuracy, we require only $O(10\%)$
accuracy in $\bar{D}\bar{w}$.   Even for maximally rotating black holes, our fits should be at least that accurate; see Fig \ref{fig:ErrorDiagnostics-GlobalVsM-Cubic}.

\begin{table}
\input{tab-mma-MomentMatrix.tex}
\caption{\label{tab:ErrorDiagnostics:MomentMatrix}Coefficient matrix $\left<\psi_a\psi_b\right>$ described in Eqs. (\ref{eq:err:J},\ref{eq:err:CoefMatrix}).
}
\end{table}

\section{NR simulations }
\label{sec:NR}

Table \ref{tab:SimulationSet} lists the waveforms used in the present work.  These waveforms were produced with
\texttt{MayaKranc}, which was used in previous \bbh{} studies \cite{2007CQGra..24...33H,Herrmann:2007ex,Herrmann:2007ac,Hinder:2007qu,Healy:2008js,Hinder:2008kv,Healy:2009zm,Healy:2009ir,Bode:2009mt}.
The grid structure for each run consisted of 10 levels of refinement 
provided by \texttt{CARPET} \cite{Schnetter-etal-03b}, a 
mesh refinement package for \texttt{CACTUS} \cite{cactus-web}. 
Sixth-order spatial finite differencing was used with the BSSN equations 
implemented with Kranc \cite{Husa:2004ip}. The outer boundaries are 
located at 317M. Each simulation was performed with a resolution of $M/77$
on the finest refinement level, with each successive level's resolution
decreased by a factor of 2.  All \bbh{} simulations have two equal-mass 
\bh{s} with total mass $M = m_1 + m_2 = 1.0$ initiated on the $x$-axis, 
with initial separation as in Table 
\ref{tab:SimulationSet}.  Table \ref{tab:SimulationSet}  also lists 
the initial spin configuration and the length of the simulation and waveform.

Because we attempt to fit to small changes in the amplitude versus spins, we carefully estimate the effects of possible
numerical artifacts, such as waveform extraction radius and simulation resolution.
For example, in a few cases waveforms were 
generated at alternate resolutions; %
convergence consistent with our fourth  order code is found.

Our best fit coefficient functions and their (fitting) uncertainties are provided in Figures
\ref{fig:Aligned:FitCoefficients} and \ref{fig:Generic:FitCoefficients}.   These fits reproduce data from our
highest-resolution simulations, extrapolated to infinite radius.

\begin{table*}
{\small
\input{tab-mma-SimulationSet}
}
\caption{\label{tab:SimulationSet} The set of equal-mass merger simulations used in this paper.  As described in Section
  \ref{sec:sub:BlindTest}, the first
  2 simulations have randomly chosen spin orientations; were not used in any fit; and provided a
  \emph{blind test} of our fitting procedure.   In this table, initial conditions are specified by the first six
  columns, which provide
  the component spins $S_k/M^2$ of each black hole, along with the initial separation $r_{start}$.  Two columns ($T,T_{wave}$) provide the duration
  of the simulation in its entirety and of the resolved, converging portion of the $l=m=2$ waveform, respectively.
  Finally, $h_{min}$ shows the smallest resolution used in our analysis.   
  All simulations have outer boundary at $r=317 M$ and radial mesh refinement boundaries including $r=79,158.7$. 
}
\end{table*}

\ROS{Request for comment!}
As seen in Table \ref{tab:SimulationSet}, we employ two choices for the initial separation: $r=10,6.2M$.   The spins and
orbit of binaries started at $r=10M$ precess, evolving into a slightly different configuration by $r=6.2 M$.  
The spin and $\hat{L}$ configuration at early times is not identical to (but can be reconstructed from) our simulations' starting
point.  Similar precession effects have been included in fits to merger recoil kicks, to correctly reconstruct the kick
direction as a function of spins at very early times \cite{gr-nr-io-fitting-BR2009,2010PhRvD..81h4054K}.
These 9 spin-misaligned systems are dominated by their aligned spins and thus precess only very slightly between
$r=10,6.2$ (e.g., $\hat{z}\cdot \hat{L}>0.96$).  However, to
avoid systematic errors caused by different starting radii, for these simulations we extract the (coordinate) spins and $\hat{L}$ at $r=6.2$.
In fact, our overall answer changes little, independent of whether $r=6.2$ or $r=10$ is used.

\section{Comparison with previous results: Aligned spin }
\label{sec:Results:AlignedSpin}

For equal-mass binaries with spins aligned with the orbital angular momentum ($P\chip=P\chim=0$), many previous studies have provided
numerical waveforms \cite{nr-Goddard-EarlyInspiralSummary-2007,gwastro-cornell-highres2008,2008PhRvD..78d4046B}, detailed tabulations of the gravitational wave amplitude $\bar{\rho}$
available to different gravitational wave detectors \cite{gwastro-nr-AlignedSpinVolumeWeight},
mismatch-based estimates of waveform complexity \cite{2007PhRvD..76h4020V,2008CQGra..25k4047S},
and even phenomenological fits to the waveforms themselves \cite{nr-Jena-nonspinning-templates2007,2008PhRvD..77j4017A,gwastro-Ajith-AlignedSpinWaveforms,2010arXiv1005.0551S}. 
To adopt a fiducial reference which directly provides comparable information, we compare our fits to \emph{generic}
spins to the \emph{spin-aligned} results provided by \citet{gwastro-nr-AlignedSpinVolumeWeight}.

\begin{figure}
\includegraphics{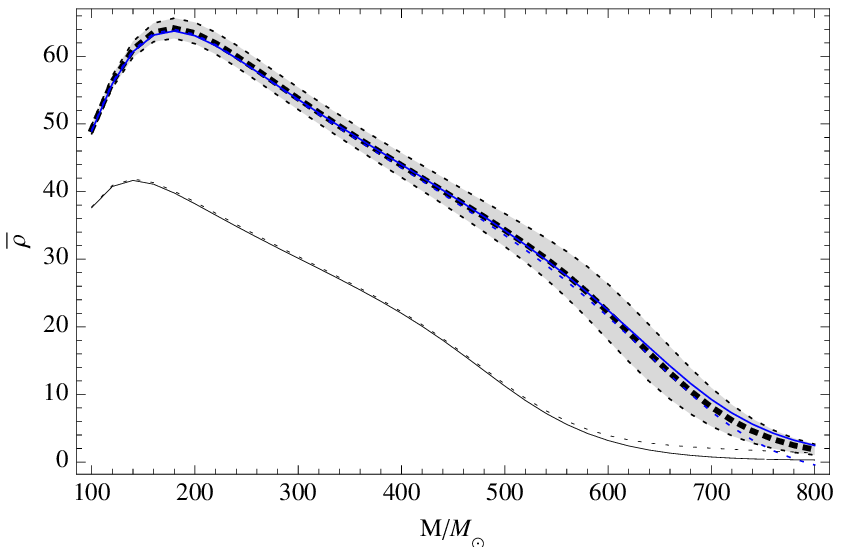}
\caption{\label{fig:Aligned:TestRange} Angle averaged
  SNR $\bar{\rho}$ (note: not $\bar{\rho}_*=\bar{\rho}/\sqrt{5}$) expected for the present LIGO
  detectors, extracted from our numerical simulations (solid lines) and from our fits (dotted lines indicate our
  preferred estimate from an unconstrained fit to all simulations; shaded interval indicates our estimate of fitting error ${\cal F}(0.8) \bar{\rho}_o$ [Eq. \ref{eq:err:J}]).
  The colored solid lines correspond to  equal-mass binaries with $a_1=a_2=0.8$ (blue, red), placed at
  $r=100\unit{Mpc}$, based on matched filtering to a single-detector datastream containing only the harmonics (i)  $l=2$
  (blue, solid); (ii)
  $l\le 4$ (blue, dotted).  For comparison, the thin black (dotted) line indicates our estimate of
  $\bar{\rho}(M)$ using $l\le 2$ ($l\le 4$) for a binary of \emph{nonspinning} black holes at the same distance. 
Compare with Fig. 6 of
  \citet{gwastro-nr-AlignedSpinVolumeWeight} and our Figure \ref{fig:Aligned:HigherHarmonics}, noting (i) $\bar{\rho}=\bar{\rho}_*\sqrt{5}$ and $\bar{\rho}_*= \rho_{*max}/(5/2)$ for the $l=2$ subspace
  (blue) and (ii) for the $l=4$ contribution they show the \emph{peak} magnitude $\rho_{*,max}$ (depending \emph{linearly} on the
  amplitude due to the $l=4$ subspace) while we show the \emph{angle-averaged} magnitude $\bar{\rho}$ (depending
  \emph{quadratically} on this amplitude and therefore highly suppressed).   Both papers adopt a comparable initial LIGO
  noise curve.    At those $M$ with a large detection volume, the SNR including higher harmonics is nearly
  indistinguishable from the SNR from $l=2$ alone.  Conversely, for sufficiently high masses (e.g., $M\gtrsim 600
  M_\odot$ for $a=0$), the $l=4$ mode dominates the angle-averaged power $\bar{\rho}$; see, for example, the two curves
  corresponding to $a=0$.
However, though $l=4$ does dominate  at the highest masses, systematic errors associated with extrapolating the
extraction radius to infinity can lead to apparent contradictions.  Here,  for $a=0.8$ the extrapolated ``total'' SNR
from all modes $l\le4$  is \emph{slightly less} than the corresponding extrapolated $\rho(M)$ curve including only
$l=2$.    \ROS{Request for comment}
}
\end{figure}

\begin{figure}
\includegraphics{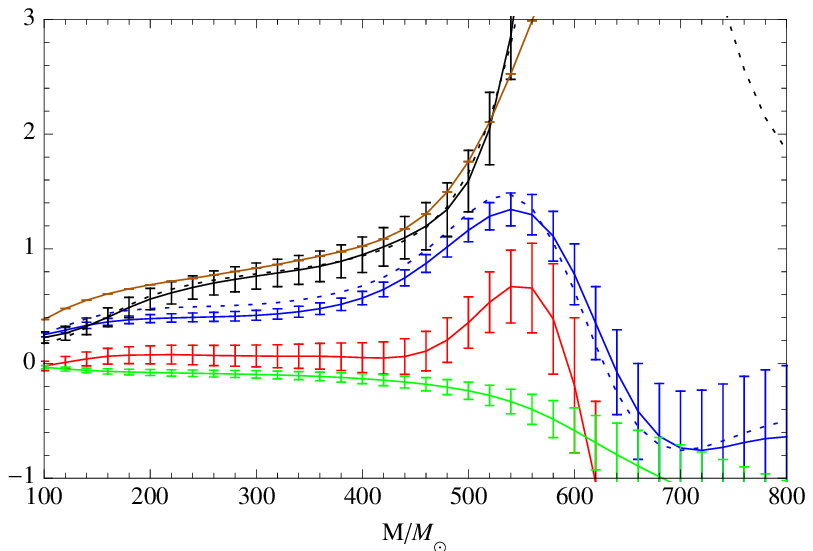}
\caption{\label{fig:Aligned:FitCoefficients} Fit coefficient functions ${\cal X}_{1,2,3}$ (blue, red, black,
  respectively), ${\cal A}_{20}$ (green), and ${\cal B}_{1020}$ (orange), along with $1\sigma$ least-squares
 \emph{fitting}  uncertainties $\sqrt{\Sigma_{kk}}$ in each parameter.  [Error bars are suppressed for the
 ill-constrained cubic-order term ${\cal B}_{1020}$ (orange).  This cubic term's  errors large and strongly
 correlated with others; also, this term's preferred fit \emph{evolves significantly} with extraction radius, even
 though each slice is consistent with ${\cal B}_{1020}\approx 0$.]
Solid lines show parameters of our general
 multiparameter fit to all available simulations;  dotted lines correspond to fits to only \emph{aligned}-spin
  data $S_{1,2}\propto \myvector{\hat{z}}$ using a special restricted model 
 only ${\cal X}_{1,3}$ are nonzero.  
}
\end{figure}

Equal-mass spin-aligned systems preserve a symmetry direction throughout their inspiral, insuring the  $l=|m|=2$ modes dominate
above higher (even) $l$ orders.
Limiting attention to the $l=2$ subspace, the \emph{peak amplitude} $\rho_{*,max}$ and \emph{source- and detector-
  orientation averaged amplitude}
$\bar{\rho}_*$ are related by\footnote{\citet{gwastro-nr-AlignedSpinVolumeWeight} provide a comparable expression,
  $\rho_{*,avg} = \sqrt{5/2} \rho_{*,max}$, relating the peak SNR $\rho_{*,max}$ to the
  \emph{source}-orientation-averaged SNR $\rho_{*,avg}$.  Their paper adopts a similar notation, without a
  single-detector $*$ subscript. }
\begin{eqnarray}
\rho_{*max}=\frac{5}{2} \bar{\rho}_* \qquad \text{l=2 only}
\end{eqnarray}
The $l>2$ modes in general and the $l=4$ modes in particular violate this relation: the signal power $\rho_{*,max}$ seen viewing along the symmetry axis increases \emph{linearly} with
$l>2$ amplitude; however, the \emph{orientation-averaged power} $\bar{\rho}$ increases \emph{quadratically} in the
higher-harmonic  amplitude (here, $l=4$).   Keeping this
distinction in mind,  [Fig. \ref{fig:Aligned:TestRange}] agree with  \citet{gwastro-nr-AlignedSpinVolumeWeight} [their
Fig. 6], who find that  for lower-mass mergers $M\le 500 M_\odot$, $l=4$ modes are typically a $5-10\%$ correction in
$\rho_{max}$ (depending on
spin) to the individual signal-to-noise ratios $\rho_{max}$, but a much smaller correction to $\bar{\rho}$ and the ratio $\bar{\rho}/\bar{\rho}_o$.
Moreover, as in all previous studies, aligned-spin waveforms appear to depend only extremely weakly if at all on the antisymmetric spin
combination ($\chim$) \cite{2008CQGra..25k4047S}, mostly through power-suppressed higher harmonics,\footnote{Previous
  studies have demonstrated that  the 2,2 mode of spin-antialigned systems resembles the nonspinning waveform.  When higher harmonics are included, the
  nonspinning and spin-antialigned waveforms have appreciable differences, measured by their relative mismatch.
  However, these higher modes carry comparatively little power: see, for example, Figure 2 of
  \citet{2008CQGra..25k4047S}, keeping in mind modes contribute power to $\bar{\rho}/\bar{\rho}_o$ in quadrature.
  Therefore, though $\bar{\rho}$ depends \emph{slightly} on the antisymmetric spin $\chim$, to a leading-order
  approximation their effect can be neglected.
}  suggesting ${\cal A}_{20}={\cal B}_{1020}\approx 0$. 
Additionally, as previous studies have shown, the orientation-averaged  amplitude $\bar{\rho}$ increases \emph{monotonically} with $\chip\cdot z$.  In terms of
our expansion,  its parameters must satisfy $\partial_{{\chip\cdot \myvector{\hat{z}}}} F >0$, or
\begin{eqnarray}
\label{eq:constraint:aligned:monotonic}
{\cal X}_1 + 2{\cal X}_2(\chip\cdot \myvector{\hat{z}}) + 3 {\cal X}_3(\chip\cdot \myvector{\hat{z}})^2 >0\,.
\end{eqnarray}
Figure  \ref{fig:Aligned:FitCoefficients}  provides  our best estimates for the coefficient functions for $\bar{\rho}$ relevant to aligned spins, both employing all
data (solid) and only $l=2$ aligned data (dashed); all satisfy this property.
Though our general fit allows for small nonzero ${\cal A}_{20},{\cal B}_{1020},{\cal X}_2$, our equal-mass aligned spin data is
well fit assuming these three parameters are exactly zero and the aligned-spin model consists of only ${\cal X}_{1,3}$
(dotted lines in Figure  \ref{fig:Aligned:FitCoefficients}; for comparison, the fitting error $\delta F$ of this
restricted model to aligned-only data is comparable to the fitting error shown in Figure
\ref{fig:ErrorDiagnostics-GlobalVsM-Cubic} for a generic model to all data).
Likewise, as our global fit error is often comparable to the change in $\bar{\rho}$ due to higher harmonics,
particularly beyond $l=4$ for $M\le 500 M_\odot$, these harmonics can to a first approximation be neglected.
Finally, for the mass range where $l=2$ emission dominates, the beampattern is extremely well approximated by the fiducial
nonspinning values.

\begin{figure}
\includegraphics[width=\columnwidth]{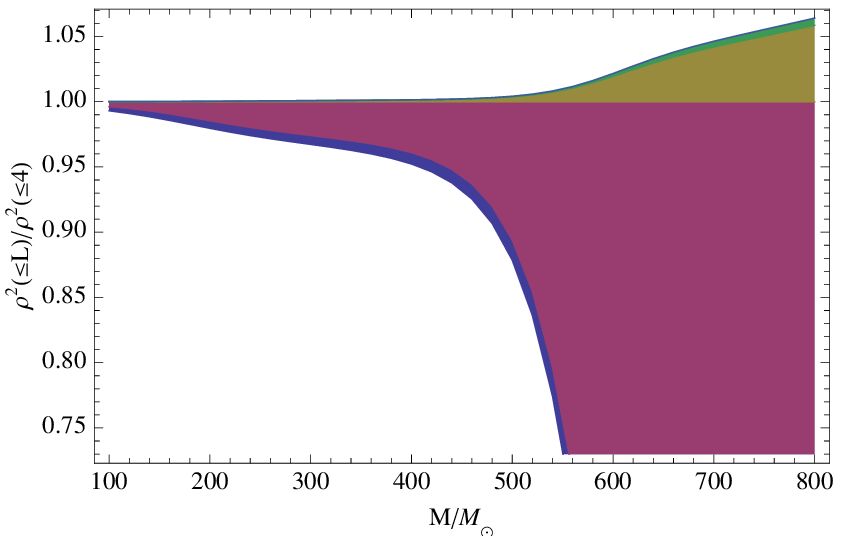}
\includegraphics[width=\columnwidth]{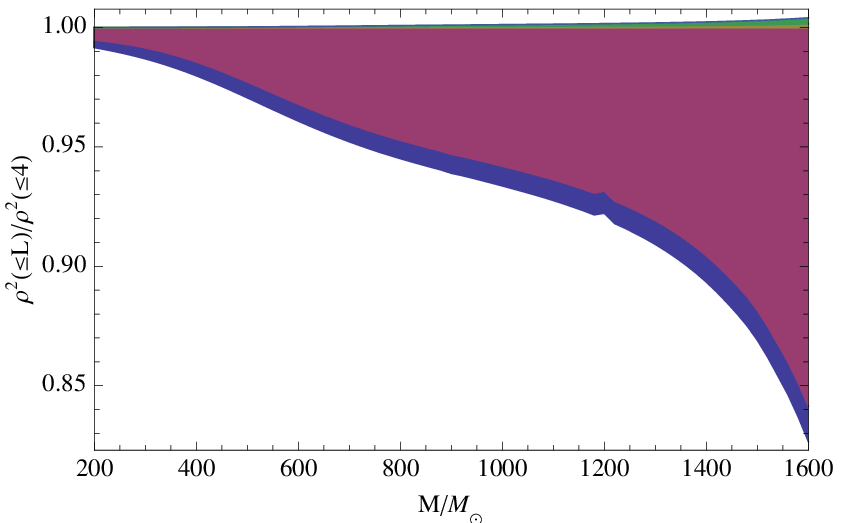}
\caption{\label{fig:Aligned:HigherHarmonics}  The relative significance of signal power due to different
  harmonics, shown for a \emph{nonspinning} equal-mass binary seen by initial LIGO (top panel) and advanced LIGO (bottom
  panel), shown at an extraction radius $r=75$.  The
  distances between the curves indicate $\bar{\rho}^2_L$, the signal power due to all $\rho$ harmonics.  At each mass,
  $\bar{\rho}^2_L$ is normalized to the contribution from all $l\le 4$ modes ($\bar{\rho}^2_{\le 4}$) 
For advanced detectors, the quadrupole mode strongly dominates for most masses.  Advanced detectors have much better low-frequency frequency sensitivity
and particularly a much less steep low-frequency limit.  On the contrary, low-mass detectors have a steep low-frequency
boundary, strongly filtering out the $l=2$ mode once the final mass and spin lead to ringdown out of band.
}
\end{figure}

For a fixed bandpass detector, high mass favors higher harmonics, particularly when those harmonics are emitted from inspiral while
 the $l=2$
modes arise from the exponentially
suppressed ringdown phase.   For aligned binaries above $M\simeq 500
M_\odot$, though the overall range drops, the $l=4$ harmonics provide an increasingly significant fraction of detectable
power, even in the absence of spin; see Figure \ref{fig:Aligned:HigherHarmonics}.  As a result,  based on the raw simulation data the beampattern
correction for initial LIGO detectors evolves from $\bar{w}_*\approx \wStarNonspinning $ at low mass (appropriate to $l=|m|=2$) down
to a relatively isotropic $\bar{w}_*\approx 1.04$ at $M\simeq 600 M_\odot$ during the transition between $l=2$ and $l=4$
beampatterns before eventually rising back at higher
masses (i.e., approaching the value $1.085$ appropriate to $l=|m|=4$; see
Table \ref{tab:CanonicalBeamingFactors} and Appendix \ref{ap:Beaming}).
In principle, the process continues with higher harmonics; in practice, however, the detection volume is sufficiently
small that their contribution to detection is negligible.

Though we estimate $\bar{\rho}$ as the nonspinning value $\bar{\rho}_o$  times a small ``correction'' $F$, at high mass and
spin the raw numerical data requires extremely large  $F$.  Comparing Figure \ref{fig:Aligned:TestRange} with 
Figure \ref{fig:ErrorDiagnostics-GlobalVsM-Cubic}, when the ratio of nonspinning to spinning $\bar{\rho}$ is large, our fit breaks down.

\subsection*{Convergence I: Extraction Radius }
\label{sec:Results:AlignedSpin:Convergence:ExtractionRadius}
\begin{figure}
\includegraphics[width=\columnwidth]{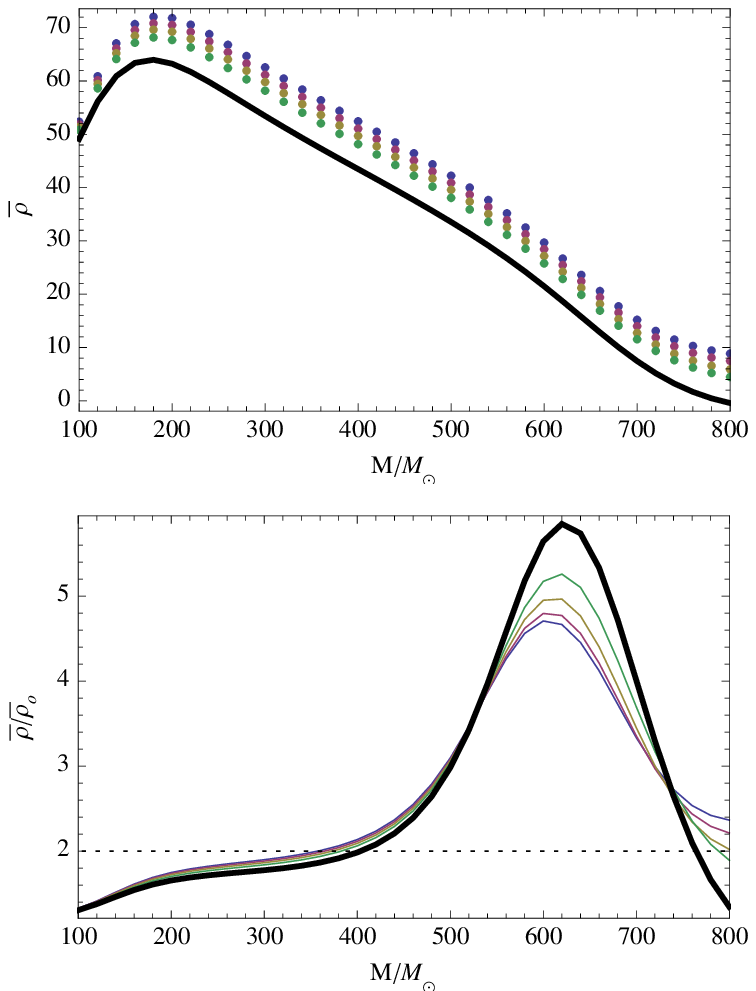}
\caption{\label{fig:AlignedSpin:Test:ExtractionRadius} For an equal-mass aligned-spin binary with $a_1=a_2=0.8$,
  estimates for 
  $\bar{\rho}$ (top panel) and  $\bar{\rho}/\bar{\rho}_o$ (bottom panel) seen by a single initial LIGO detector at
  $D=100\unit{Mpc}$ versus total binary mass $M$, based on
  waveforms extracted from our numerical merger simulations at 
  coordinate extraction radii
  $r=40,50,60,75$  (blue, red, yellow, green), and (thick solid) linearly 
  extrapolated in $1/r$ to $r\rightarrow \infty$.   For high-mass mergers $M\ge 500 M_\odot$,  the aligned-spin
  range is much larger than the nonspinning range 
  ($\bar{\rho}/\bar{\rho}_o>2$; compare to Figure \ref{fig:Aligned:TestRange}).   We anticipate both our extrapolation
  to large radius and our fit to perform poorly when the ratio $\bar{\rho}/\bar{\rho}_o$ is large.
}
\end{figure}

Our low-resolution large-radius grid zones do not retain enough information to permit adequately accurate waveform
extraction for all harmonics.  Therefore, unlike 
  \citet{gwastro-nr-AlignedSpinVolumeWeight} who extract at $r=160 M$ and particularly unlike extraction at ${\cal J}^+$
  \cite{gr-nr-methods-ExtractionAtInfinity-Reisswig2009}, we extract relatively close to the binary,
  at coordinate
  radii $r= 40,50,60$ (and, when available, at $r=75$).   As seen in Figure \ref{fig:AlignedSpin:Test:ExtractionRadius}, finite extraction radius effects  can compete with the small spin-dependent
  changes in $\bar{\rho}$.

For this reason, rather than adopt a single preferred extraction radius, we first extrapolate $\rho(M)$ to
infinity, then fit to the extrapolated $\rho(M)$ data. 
Owing to instabilities in the extrapolation at high mass to $\bar{\rho}$ calculated using initial detector noise spectrum $S_h$, we do not trust our fit for extremely high masses $M\ge 500
M_\odot$.  On the contrary, for advanced detectors which lack such a steep low-frequency cutoff, our fitting
procedure works well to proportionally higher masses; see Appendix \ref{ap:AdvancedDetectorFits}

\begin{figure}
\includegraphics[width=\columnwidth]{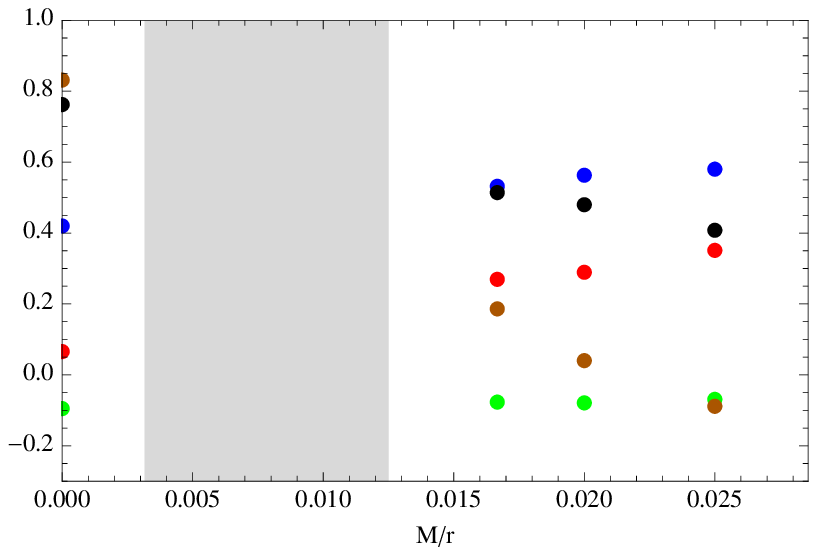}
\caption{\label{fig:AlignedSpin:Test:ExtractionRadius:Parameters} For an equal-mass binary with $M=300 M_\odot$ and
  generic spins,
  estimates for the fitting parameters ${\cal X}_{1,2,3}$ and ${\cal B}_{1020}$ to $\bar{\rho}/\bar{\rho}_o$, based on
  numerical waveforms extracted on spheres of radius  $r=40,50,60$, shown   versus $1/r$.  Colors and symbols are associated as
in Figure \ref{fig:Aligned:FitCoefficients}.  This figure provides neither point error estimates nor \emph{extremely
  strong correlations} between recovered parameters.   Also shown at $1/r=0$ are the best-fit
  parameters to $\bar{\rho}/\bar{\rho}_o$ data that has been linearly extrapolated to $r\rightarrow \infty$ based on these radial slices
 as well as to the often-available $r=75$ slice.  Finally, the light gray region indicates the range
  of extraction radii associated with our two lowest-resolution zones (i.e., between $r=80$ and $317$); to avoid
  introducing systematic error, we extrapolate over radii associated with a single grid zone.
}
\end{figure}

Figure \ref{fig:AlignedSpin:Test:ExtractionRadius:Parameters} also  demonstrates that specific coefficient values are  sensitive to extrapolation, particularly the
high-order (cubic) coefficients.
This trend versus extraction radius demonstrates a key issue associated with any phenomenological fit:  while we can
present best-fit parameter estimates, these estimates are subject to strong correlations.

Despite uncertain and highly correlated fitting \emph{parameters}, the best fit \emph{functions} extracted from each extraction radius and
from the radially extrapolated data are both extremely consistent with one another and with decreasing differences as
$r$ increases.  Specifically, performing Monte Carlo estimates of the $L^2$ difference between the
finite-radius fits and the fit to extrapolated data, we find the set of $L^2$ functional differences $||F_{r}-F_{\infty}||$
(i)  is consistent with being proportional to $1/r$ and (ii) at each radius is comparable to
the semianalytic error estimate shown in Fig. \ref{fig:ErrorDiagnostics-GlobalVsM-Cubic}.

\section{Selection biases for generic, comparable-mass mergers}
\label{sec:Results:Generic}

Even well before the epochs considered here, binaries with generic spins precess, breaking symmetry and distributing
power among other $l=2$ modes.   The asymmetric merger process and ringdown   favors exciting otherwise-suppressed higher harmonics
(e.g., $l=3$), as well as the $l=4$ modes present even without spin.   If the merger is in band, corresponding to masses $M\in[50,500]M_\odot$, higher
harmonics contribute an increasing proportion of overall SNR, particularly at high spin; see, e.g., Fig. 2 in
\citet{2008CQGra..25k4047S}.   As in the nonspinning case,  if late stages of merger are detected, corresponding to masses $M\gtrsim 500 M_\odot$ for initial and
$M\gtrsim 1000 M_\odot$ for advanced detectors, the beampattern can be significantly asymmetric.
Unfortunately, though substantial beampattern asymmetries begin to occur at high mass and spin ($\bar{w}_*(M)$ varying), our fit to
the angle-averaged power  ($\bar{\rho}/\bar{\rho}_o$) is not accurate enough at these high masses to justify a detailed
analysis.  At lower mass, spin precession only weakly anisotropizes the beam.  Comparing to our symmetry expansion
[Eq. (\ref{eq:expansion:barrho})], any (linear) leading-order spin-dependent corrections depend on only \emph{aligned}
spin components; the linear term is without loss of generality determined by simulations of aligned merging binaries.
As the beampatterns from aligned merging binaries are dominated by  $l=|m|=2$ emission,
the beampattern correction factor for \emph{generic misaligned binaries}  $\bar{w}_*$ changes little from its nonspinning value.  
Though the beampattern changes \emph{shape} substantially due to precession, its \emph{effective volume} is nearly unchanged.

\begin{figure}
\includegraphics[width=\columnwidth]{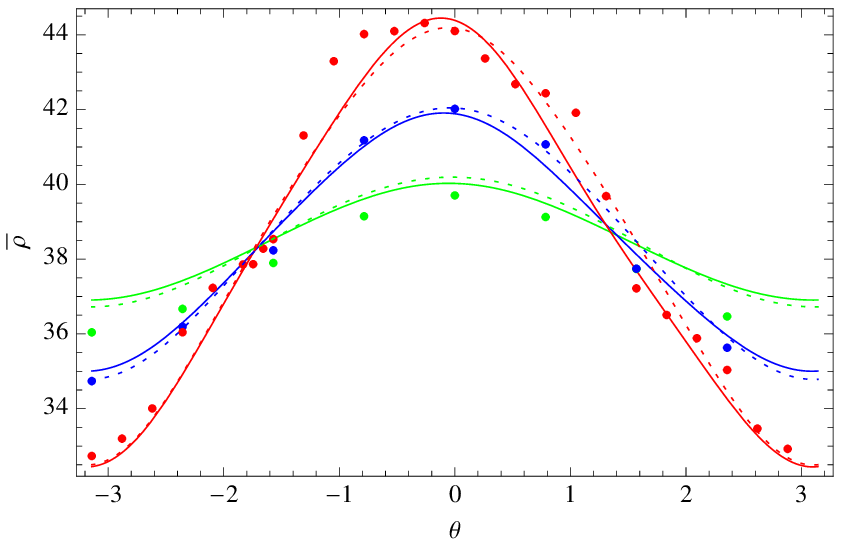}
\caption{\label{fig:SSeries:FversusFit} For a $M=200M_\odot$ binary, angle-averaged SNR ratio
  $F=\bar{\rho}/\bar{\rho}_o$ (points); our generic, unconstrained cubic-order fit
  $\hat{F}$ to all simulations (solid line); and a fit where only ${\cal X}_{1,2,3}$ and ${\cal X}_{02}$ can be nonzero
  (dashed lines).    The data and
  fits are shown  for spin magnitudes $a=0.4,0.6,0.8$ (green, blue, red, respectively) and spins
  $\myvector{S}_1$ in the orbital plane (along the separation vector) and $\myvector{S}_2$ tilted by angle $\theta$ away from $\myvector{\hat{z}}$.
}\end{figure}

\begin{figure}
\includegraphics[width=\columnwidth]{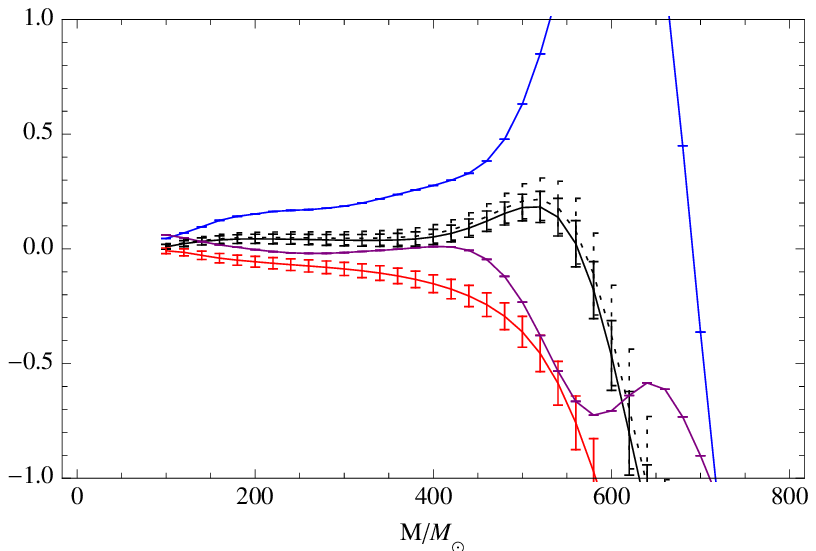}
\caption{\label{fig:Generic:FitCoefficients} Fit coefficient functions ${\cal X}_{02},{\cal A}_{02},{\cal
    B}_{1200,1002}$ (black, red, blue,purple) and estimated least-squares fitting
  uncertainties $\Sigma_{kk}$ for the coefficients relevant to misaligned spin; compare with Figure
  \ref{fig:Aligned:FitCoefficients}.   As in that figure, the large fitting errors ($\gtrsim 0.25$) on cubic-order terms (${\cal
    B}_{1200}, {\cal B}_{1002}$) are not shown.  Solid lines show a general fit including all harmonics and spins.  The dotted black
  line corresponds to the value of ${\cal X}_{02}$ assuming a model where only it and ${\cal X}_{1,2,3}$ are nonzero.
}
\end{figure}

On the other hand, the average amplitude $\bar{\rho}$ varies substantially with spin magnitude and orientation.
Well-known results for aligned spins illustrate the  relative impact that spin orientations can produce.  As a function
of spin orientations relative to $\hat{L}$,   the largest $\bar{\rho}$ occur with spins that are aligned with the orbit  ($\chim=0;
\myvector{S}_{1,2}\propto \myvector{\hat{z}}$);  the smallest $\bar{\rho}$ occurs with spins that are antialigned with the orbit
($\chim=0;S_{1,2}\propto -\myvector{\hat{z}}$); and $\bar{\rho}$ is essentially unchanged if the spins are mutually antialigned
($\chip=0$, both for $S_{1,2}\propto \myvector{\hat{z}}$ and generally \emph{for all spin orientations} $\chim$). 
In fact, for these these intuitively obvious conditions to hold, the expansion parameters in
Eq. (\ref{eq:expansion:barrho}) must satisfy the following conditions:
\begin{eqnarray}
{\cal A}_{20}\approx {\cal A}_{02}&\approx& 0 \nonumber \\
{\cal X}_1 \pm |\chip| {\cal X}_2 + \frac{3}{4}|\chip|^2 {\cal X}_3& >&  \frac{{\cal B}_{1200}|\chip|^2}{2}  \pm
|\chip|{\cal X}_{02} \nonumber
\end{eqnarray}
all of which our best-fit coefficients satisfy.   As noted previously, because we included cubic-order terms, our fit for $\bar{\rho}$ is also
consistent with $\bar{\rho}$ being a positive and monotonically increasing function of equal, aligned spins; see Eq. (\ref{eq:constraint:aligned:monotonic}).
More generally, for arbitrary spin orientations our fit is positive-definite for at least $M<500 M_\odot$ and $|a_{1,2}|<0.8$, as well as for selected spin
configurations at higher mass.
As in the nonspinning case, however, our fit breaks down for binaries with large aligned spins and high mass, as in
these regions 
the correction factor $F=\bar{\rho}/\bar{\rho}_o$ must be nearly $0$ (for spins mostly antialigned with the orbit) or much
larger than unity (for component spins mostly aligned with the orbit).
Figures \ref{fig:Aligned:FitCoefficients} and  \ref{fig:Generic:FitCoefficients} provide our best-fit coefficients to the expansion of
Eq. (\ref{eq:expansion:barrho}).  These preferred values reproduce our best simulation resolutions,
extrapolated to $r\rightarrow \infty$, including all available harmonics.\footnote{For completeness, in the online version of this
article we provide data for and fits to the cases where only $l\le 2,3,4,5$ harmonics were used.  Except for low spin,
fitting errors are comparable to or larger than the contribution of higher harmonics.} 
Though the symmetry expansion of Eq. (\ref{eq:expansion:barrho}) permits  more generic behavior with spin, our numerical results
are well-fit with a far more restrictive form where only ${\cal X}_{1,2,3}$ and ${\cal X}_{02}$ are nonzero.\footnote{In
  fact, roughly speaking this restricted fits' coefficients are themselves related by a nearly mass-independent proportionality: ${\cal X}_1\approx {\cal X}_3\approx O(2-4) {\cal X}_2 \approx
  O(6-8){\cal X}_{02}$, both for initial ($M\in[100,500]M_\odot$) and advanced ($M\in[200, 1300]M_\odot$) LIGO
  detectors.}  As shown by the bottom panel in Figure \ref{fig:ErrorDiagnostics-GlobalVsM-Cubic}, this fit
performs well \emph{averaged over all simulations}  because our simulations mostly have  $|a_{1,2}|\le 0.6$ [Table
\ref{tab:SimulationSet}], where cubic order corrections are still small.   At the same time, the fit has more than
enough parameters to explain the predominantly linear- and quadratic-order  variation in $\bar{\rho}$ versus generic spin orientations; see for example Figure \ref{fig:SSeries:FversusFit}.

\ForInternalReference{
* key points:

 - we have an accurate fit for rho, including all relevant modes.  

     : Reality check: iLIGO has problems with l=4-ish above 600 Msun.  Not a problem for aLIGO

 - our prediction is \textbf{positive-definite}

 - our prediction is more or less equivalent to a symmetry-restricted case with only X1,X2,X3, X02 nonzero.  This case
 is completely independent of chiminus.  The gradient $\nabla_{\chi_+}\rho$ is   .... does it have any points with $\nabla \rho=0$? (i.e., special points)
}

\ForInternalReference{\begin{figure}
\caption{\label{fit:Generic:BeamingCorrectionFactor:FitCoefficients} Fit coefficient ${\cal Z}_1$ in the beampattern
  correction function -- should be nearly zero, as only PERPENDICULAR precession \richardBugFix \ROS{No reason to
    include parameters of a fit to this function, given  $\bar{\rho}$ accuracy.  We barely even need its leading-order value.}
}
\end{figure}
}

\begin{figure}
\includegraphics[width=\columnwidth]{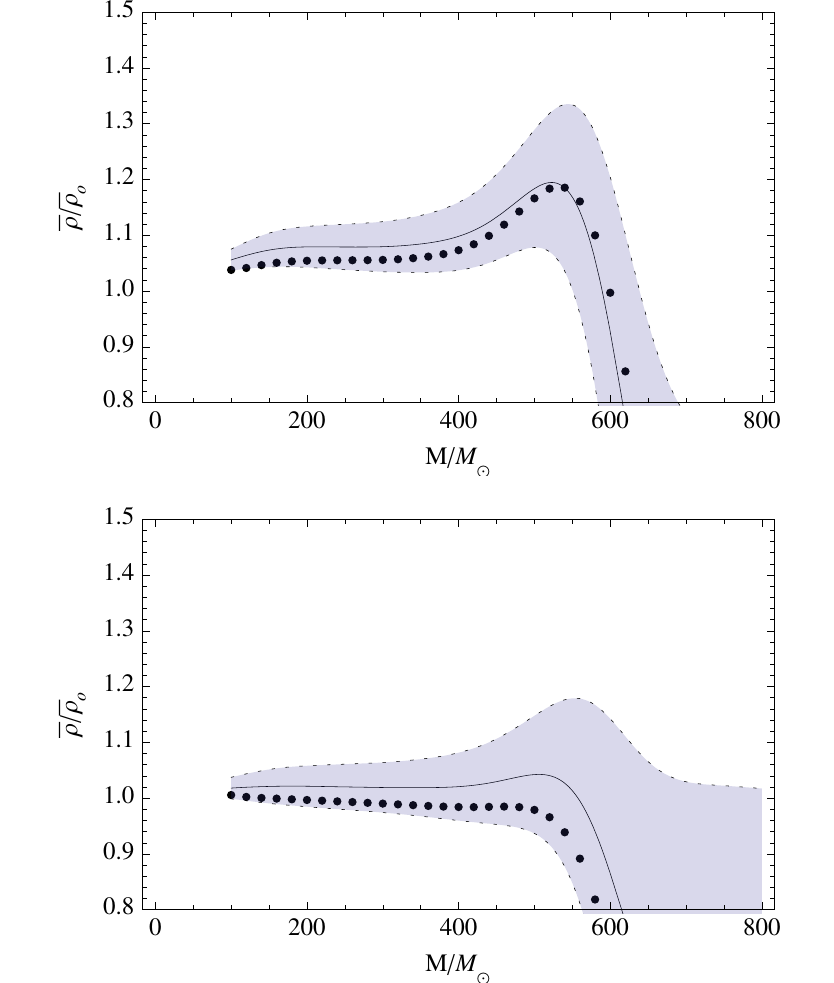}
\caption{\label{fig:Generic:BlindTest1} A plot of $F=\bar{\rho}/\bar{\rho}_o$ for the ``blind test'' spin
  configurations.  Dots show the raw results from our numerical simulation; the curves and shaded region indicate results
  from our preferred fit. 
}
\end{figure}

\noindent \emph{Blind test}
\label{sec:sub:BlindTest}
The first two simulations in Table \ref{tab:SimulationSet}, with random spins with $|a_1|=|a_2|=0.6$, were reserved as a blind
test.  As illustrated  by Figure \ref{fig:Generic:BlindTest1}, for low masses $M\le 500M_\odot$   our fit correctly recovers
$\bar{\rho}/\bar{\rho}_o$ to within roughly the $1\sigma$  relative error ${\cal F}$, shown as the shaded region on this
plot.

\ForInternalReference{
\subsection{Symmetry breaking and higher-order modes}
Previous studies have suggested the $l=|m|=2$ waveforms are surprisingly simple:  not only do they depend weakly on
antisymmetric spin components, but (being the dominant mode) their response depends only weakly on the perpendicular
spin.
In terms of our expansion, if only $l=2$ modes are included we expect at quadratic order that ${\cal A}_{20}\approx {\cal A}_{02}\approx {\cal X}_{02}\approx 0$.
Higher harmonics break this strong symmetry.  At low mass, when the $l=2$ mode still dominates, the signal amplitude
from higher harmonics
adds in quadrature and therefore only very weakly changes $\bar{\rho}$, at a level typically comparable to our fitting uncertainty.  However, at the high masses where the initial
LIGO detector strongly filters out  $l=2$ modes, the  $l=4$ modes dominate the response and therefore the fit

\begin{figure}
\includegraphics{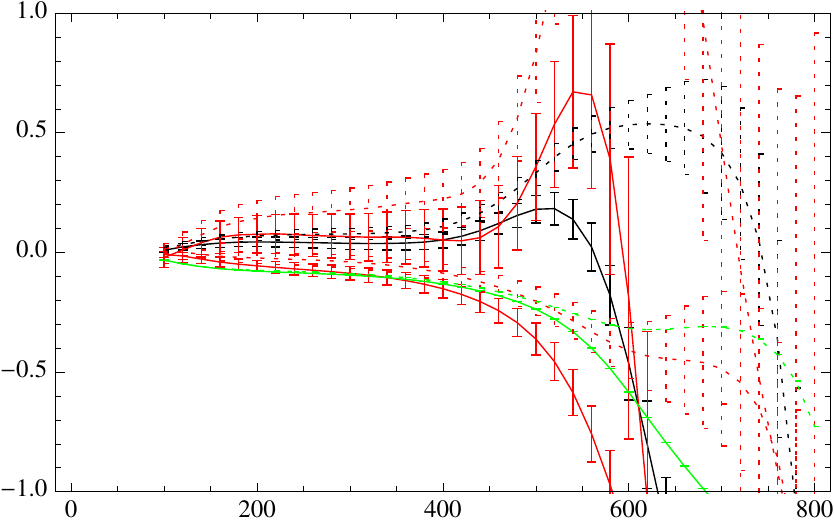}
\caption{Comparing parameters obtained using only $l=2$ and when including higher order modes.  Differences
  at large $M$ are apparent.}
\end{figure}
}

\optional{\subsection{Model stability (*)}

* how the functions vary from each other in $L^2$ : how important are those high-order correction at high spin

* Will not be included in draft version; perhaps later
}

\section{Astrophysics }
\label{sec:Astro}

Using the fits provided in this paper, we can use Eq. (\ref{eq:DetectionRate}) to determine the relative likelihood of
\emph{observing} different spin magnitudes and alignments, given a progenitor parameter distribution $p(\bar{\lambda})$.   As expected from studies of
aligned spins, ground-based gravitational wave detectors will be  biased towards the detection of aligned and (to a lesser
extent) large spins.  For example, as seen in Figure \ref{fig:astro:DetectionVolumeVersusSpin} in the special case
$M=200 M_\odot$, populations of randomly
oriented spins with identical magnitude $|a_1|=|a_2|=a$ are much less likely to be seen than a
comparable population of spin-orbit aligned spins, and conversely for spin-orbit antialigned spins $a_1=a_2=-a\myvector{\hat{z}}$.
However,  populations of binaries with random spins (dotted line in Figure \ref{fig:astro:DetectionVolumeVersusSpin}) have comparable average detection volumes as \emph{nonspinning}
binaries, with detection volume increasing only $O(50\%)$ if spin magnitudes as large as $0.8$ are allowed, only
marginally larger than fitting error in the detection volume.
In other words, in a population of random spins,  any pair of spin magnitudes are roughly  likely to be observed,
with only a slight bias towards larger spin magnitudes.   Despite the much larger detection volume for aligned spins,
\emph{a priori} perfectly aligned and large spins should be rare, suppressing the bias towards high spin magnitude.
For similar reasons, in any population of \emph{random} spin orientations,  large aligned spins occur rarely enough and
the bias towards large spin is small enough (typically less than a factor 5) that
the associated  detected population will not be overabundant in tightly aligned spins unless the progenitor population
is.  For the purposes of illustration, suppose we  assume any spin pair with both spins $S_{1,2}$
within  $\pi/4$ of alignment with $\myvector{\hat{z}}$ (i.e., $\hat{S}_{1,2}\cdot\myvector{\hat{z}}>1/\sqrt{2}$) will be significantly amplified, independent of spin magnitude.   Even in this
unrealistically optimistic case,  only $[(1-\cos\pi/4)/2]^2\simeq 2\%$ of
all random progenitor spin orientations could be amplified.

All the properties described above and exhibited in Figure \ref{fig:astro:DetectionVolumeVersusSpin} can be understood
analytically.  For simplicity, let us adopt a fit with only ${\cal X}_{1,2,3}$ and ${\cal X}_{02}$ nonzero (dotted lines
in Figures \ref{fig:Aligned:FitCoefficients},\ref{fig:Generic:FitCoefficients}).  The detection volume can be approximated to
quadratic order as 
\begin{eqnarray}
V(\myvector{S}_1,\myvector{S}_2)&\propto& \bar{\rho}_0^3[1+3{\cal X}_1 (\chip\cdot\myvector{\hat{z}})
  + 3({\cal X}_1^2+{\cal X}_2)(\chip\cdot \myvector{\hat{z}})^2 
  \nonumber \\ &+& 3{\cal X}_{02}(P\chip)^2 \,.
  + O(\chi^3)] 
\end{eqnarray}
Averaging the detection volume over all spin orientations (or any symmetric volumes in $\myvector{S}_1,\myvector{S}_2$) eliminates terms of
odd order, leaving only quadratic-order dependence on spin:\footnote{For spins with identical spin magnitude
  $|a_{1,2}|=a$, the angle-averaged coefficients of this expansion are provided by Table \ref{tab:ErrorDiagnostics:MomentMatrix}.}
\begin{eqnarray}
\left<V(\myvector{S}_1,\myvector{S}_2)\right>_{\Omega}& \propto& \bar{\rho}_0^3[1
  + 3({\cal X}_1^2+{\cal X}_2)\left<(\chip\cdot \myvector{\hat{z}})^2\right>_\Omega
  \nonumber \\ &+& 3{\cal X}_{02}\left<(P\chip)^2 \right>_\Omega
 + O(\chi)^4]\, .
\end{eqnarray}
This expression reproduces the corresponding 
(blue dotted) curves in Fig. \ref{fig:astro:DetectionVolumeVersusSpin}.
Strictly speaking, our simulations and fit apply only to equal mass ratio.  However, given symmetry considerations, the
leading-order spin- and $\delta m/M$-dependent corrections to the detection volume must be \emph{quadratic} in the
asymmetric mass ratio times a quadratic function of spins.  Our estimate therefore applies to many comparable-mass IMBH-IMBH mergers as well.

\begin{figure}
\includegraphics[width=\columnwidth]{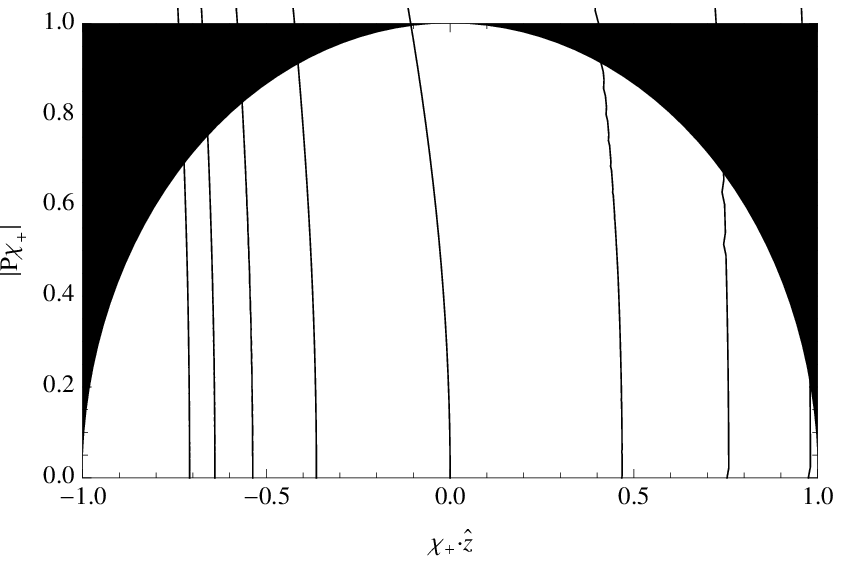}
\caption{\label{fig:Astro:DetectionVolumeVersusSpinContours} Contours of single-detector initial LIGO detection volume
  to an equal-mass $m_1=m_2=200 M_\odot$ binary versus the symmetric spin
  $\chip$, adopting a simple fit with only
  ${\cal X}_{1,2,3}$ and ${\cal X}_{02}$ nonzero.    Contours are shown for $V=2^n V_o$ for $n=-4,-3\ldots 3$, increasing
  from left to right on the graph, where $V_o$ is the detection volume for zero spin.  The shaded regions indicate
  physically unattainable $\chip$ values (i.e., which must violate  $|a_{1,2}|\le 1$).
}
\end{figure}

\begin{figure}
\includegraphics{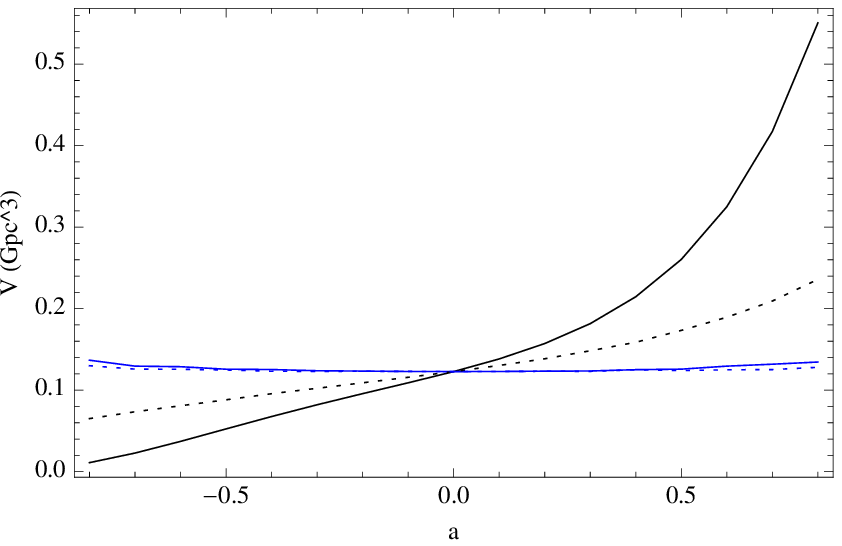}
\includegraphics[width=\columnwidth]{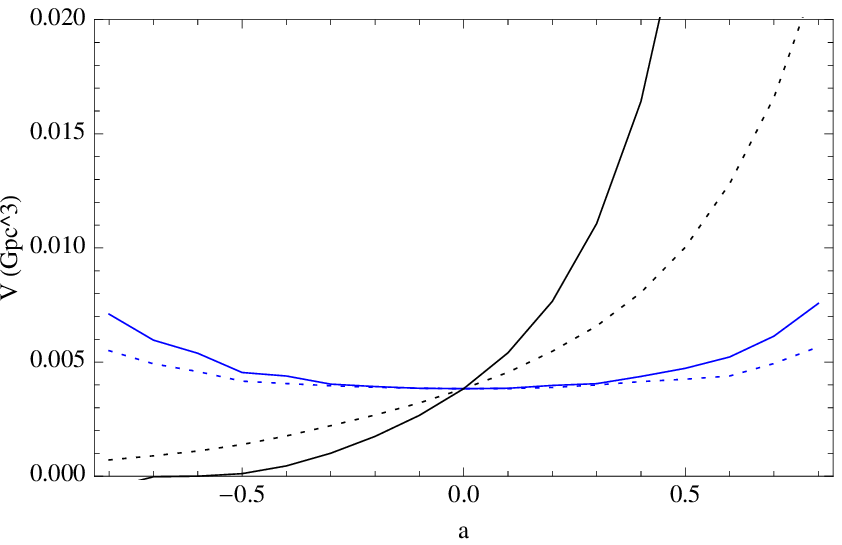}
\caption{\label{fig:astro:DetectionVolumeVersusSpin}
Detection volume for initial LIGO for BH-BH binary populations with $m_1=m_2=M/2=100M_\odot$ and different spin orientation and magnitude
distributions:  aligned, identical spins (black, solid); aligned spins independently distributed uniformly between $[0,a]$ or
$[-|a|,0]$, if $a<0$
(black,dotted); randomly aligned spins with identical magnitude (blue, solid); and component spins independently drawn from a
sphere of radius $a$ (blue, dashed).
This figure provides only the single-detector beaming-uncorrected detection volume $V\equiv 4\pi (\bar{D})^3/3$ assuming
a single-detector SNR threshold $\rho_c=8$; beaming
increases the detection volume by roughly $(\wStarNonspinning)^3$.  At low masses, far from the sharp initial LIGO low-frequency
cutoff, all equal-mass binaries have a comparable detection volume. 
Bottom panel: as top panel, but now $m_1=m_2=250 M_\odot$.
}
\end{figure}

Some BH-BH binary  populations have  an intrinsic bias towards  aligned spin, such as the products of binary  evolution of extremely
low-metallicity binary black holes.  In these cases, spin-orbit misalignment encodes otherwise inaccessible information
about the strength of supernova kicks on those very massive black holes.\footnote{Very massive BHs should have supernova kicks
  strongly suppressed by fallback, or even prompt collapse.  The complete absence of spin-orbit misalignment would
  confirm this model.}   For an intrinsically aligned population, gravitational wave detectors are far more likely to
observe large, tightly aligned spins.      Figure \ref{fig:Astro:DetectionVolumeVersusSpinContours} illustrates this effect using 
contours of the
detection volume  versus $\chip\cdot\myvector{\hat{z}}=(S_{1,z}+S_{2,z})/2$ (the aligned component of spin) and $(P\chip)^2$ (the
perpendicular part of the symmetrized spin), adopting a fit with only ${\cal X}_{1,2,3}$ and ${\cal X}_{20}$ nonzero.

Below roughly $40\unit{Hz}$, the initial LIGO detector's design sensitivity to merger waves ($O(f^{-7/3}/S_h)$) decreases rapidly.  Some mergers yield a
final-state black hole whose
ringdown frequency is comparable to this frequency.  As a result, small changes in mass or spin lead to large changes in
the volume to which these black holes can be seen.  In particular, for black holes of fixed mass, this low-frequency
cutoff produces a noticable bias towards large, aligned spins.   Unfortunately, our fits to $\bar{\rho}$ degrade
precisely where this selection bias becomes important.  However, as seen in the bottom panel of Figure
\ref{fig:astro:DetectionVolumeVersusSpin}, the bias towards large spins is already apparent at $M=500 M_\odot$.
For example, 
at $a=0.8$, the blue solid line (randomly oriented spins) is twice as large as the nonspinning value, in contrast to
only tens of percent higher at lower mass.  
At this and higher masses, the distance to which a nonspinning binary is visible can be substantially greater, up to a
factor of $6$ or more near $M=600 M_\odot$; see  Fig. \ref{fig:AlignedSpin:Test:ExtractionRadius}.  This substantial
increase in detection volume $\propto (\bar{\rho}/\bar{\rho}_o)^3\simeq O(10-100)$ can partially compensate for the
rarity with which random spin directions find themselves aligned.
However, as this unusual selection bias for high spin operates only in a narrow mass and spin range and only for the initial
detectors, we do not estimate this bias to an astrophysically relevant level of accuracy.

\optional{
* interpreting population spin distribution: much more likely to see aligned spins...but the sensitivity doesn't
increase so rapidly that they necessarily \emph{dominate} the detected population, unless spins are bounded from below
(i.e., more similar to the blue curves)
}

\optional{* bias is extremely strong: a factor of up to  5 (at $|a_1|=|a_2|=0.8$, over all of $M=200 M_\odot$ to $M\approx 500 M_\odot$) for aligned spins at
peak, by iLIGO, over generic randomly-oriented spins.}

\section{Conclusions}

We have provided simple phenomenological fits to the average detection volume within which  a merging equal-mass BH-BH binary with
$M\in[100,500]M_\odot$ and \emph{arbitrary spins} would produce a signal-to-noise ratio $\rho$ greater than a fixed
detection threshold, for  a single initial  LIGO
interferometer operating at design sensitivity.
Comparable  results for advanced LIGO are described in the Appendix.
Though we describe fits for a single detector, our method also applies to arbitrary  networks of identical detectors, adopting the
universal angle-averaged power $\bar{\rho}$ combined with a
suitably-modified (and network-topology-dependent) beaming correction factor $\bar{w}$.
Our generic expression for the range is both surprisingly simple and accurate, involving at most 10 (=9+1) functions
to reproduce the response to generic spin magnitudes and orientations to cubic order in the component spins.  Moreover, we find an even simpler expression
involving only four mass-dependent terms (${\cal X}_{1,2,3}$ and ${\cal X}_{02}$) also fits all our results.  
Though derived for equal mass ratio, symmetry considerations suggest unequal mass ratio corrections enter weakly, at
higher order (in $\propto \delta m\, \chim$).  Our spin-dependent expressions for $\bar{\rho}/\bar{\rho}_o$  should
therefore correctly estimate that ratio for  all comparable-mass binaries.   Finally, our method is easily extended to
unequal mass: a \emph{three}-parameter symmetry-constrained  expansion of $\bar{\rho}/\bar{\rho}_o$ (in $\chi_\pm$ and $\delta
m/M$) still has comparatively few parameters.

The dynamical processes that most likely  produce merging BH-BH binaries at these masses likely guarantee random spin orientation
\cite{imbhlisa-2006}.
Though gravitational wave detectors are far more sensitive to BH-BH binaries with aligned spins, we conclude that BH-BH
populations with \emph{random} spin orientations will rarely provide detections from tightly aligned, high-mass BH-BH
binaries, under the assumption of optimal signal processing.
Lacking the waveforms needed to perform optimal signal processing in the high-mass region, however, present-day all-sky gravitational wave
searches conduct searches using approximate or hierarchical methods.  Further study is needed to assess how strongly
the search methodologies themselves introduce bias towards aligned spin.

Our fit suggests a surprising and astrophysically convenient conclusion: for $M\in[100,500] M_\odot$ for initial LIGO
and over $M\in[200, 1600]M_\odot$ for advanced LIGO, \optional{\editremark{confirm at high mass}} the population-averaged detection volume for 
binaries with random spin directions and an  \emph{arbitrary spin magnitude distribution} and $a_{1,2}<0.8$  is nearly
identical (within tens of percent, comparable to the Poisson error in 10 detections) to the
detection volume for \emph{nonspinning binaries of comparable mass}. 
On the contrary, detectors like initial LIGO which have both a shallow optimally sensitive region combined with a very steep low frequency cutoffs will be noticably more likely
to recover large aligned spins from a random spin population,
albeit only for the very highest masses and spins to which the detector is sensitive ($M>500 M_\odot$ for initial LIGO).

\ROS{Comment on paragraph below}
Conversely, our analysis implies that the \emph{detected} population of high-mass binaries with a given (optimally
filtered) SNR $\rho$ will be distributed uniformly over spin orientations, to the extent that their formation processes produce them.   Our study therefore reaffirms the urgent
need for models for the merger waveforms from  generic  spinning merging binaries.
At present, our fit breaks down at very high mass, empirically when  $(\bar{\rho}/\bar{\rho}_o-1)$ is of order unity.
Several avenues of improvement could make the fit more stable:  fitting to a logarithm $\ln \bar{\rho}/\bar{\rho}_o$, or
even using  similarity transformations to rescale $\bar{\rho}(M)$ to different spin geometries (e.g., changing the mass and amplitude scale using the ringdown frequency of the
post-merger BH).
We have also not compared our fitting parameters with  comparable coefficients for better-understood low-mass  binaries (e.g., those undergoing simple
precession), which can be estimated analytically and numerically.
We  will address these refinements in a future paper.

For simplicity, our analysis is expressed in terms of matched filtering and  a fixed SNR detection threshold.  Whether
due to incomplete signal models, inefficiencies in template placement, or the lack of a signal model altogether, realistic searches cannot identify all
available signal power.  Additionally,  whether from higher mass or antialigned spin, waveforms of short duration are
more easily confused with nongaussian detector noise and require a significantly higher detection threshold.
Though a significant technical challenge,  the search- and detector-dependent effective detection threshold versus binary parameters $\rho_c(\lambda)$ that
incorporates both mismatch and noise effects can also be tabulated.   For example, for low mass binaries the mismatch
between nonspinning search templates and precessing, spinning waveforms has been tabulated; see Brown et al. (in preparation).   Combined with our fit to the intrinsic available
detection volume, a ``detector sensitivity'' fit could  efficiently communicate an adequately-accurate representation of the
parameter-dependent detection volume of real high-mass searches.

To summarize, we have provided the first phenomenological fit to the spin-dependent detection horizon for generic spin
magnitudes, spin orientations,
and (equal) component masses $M\in[100,800 M_\odot]$.  
We have shown that to leading order spin effects average out of the detection volume.   For example, the
rate at which IMBH-IMBH mergers will be detected is directly proportional to this volume.   All previous estimates for the
IMBH-IMBH detection rate estimated this volume assuming minimal or occasionally aligned component spins
\cite{gw-astro-ET-IMBH-ReviewMiller-2009,gwastro-imbh-ComparableMergers-2009,gwastro-imbh-ComparableMergers-Ilya-2009}
and therefore should be nearly unchanged even with spin included.

\optional{Comments from LIGO people: 

- Fig 11 is discussed in the paper before fig 10 : Wait for editorial staff at journal

- add refs to Ilya, on extreme mass ratio: 0705.0285v2 \cite{gw-astro-imbh-imri-2009}

- explain transition to $l=4$ increasing at 1200 Msun, possible explanation for qualitative changes starting in fits

}

\begin{acknowledgements}
 ROS is
supported by NSF award PHY 06-53462, PHY-0970074, the Center for  Gravitational Wave Physics, and the Center for
Gravitation and Cosmology.  BV is supported by the Center for
Gravitational Wave Astronomy under NSF CREST 0734800.  This work is also supported by NSF grants to DS PHY-0925345, PHY-0941417, PHY-0903973 and TG-PHY060013N.  
 We also thank Tanja Bode, Frank Herrmann,
Ian Hinder and Pablo Laguna for their contributions to the \texttt{MayaKranc} code.
\end{acknowledgements}

\appendix

\section{Selection biases of individual advanced detectors }
\label{ap:AdvancedDetectorFits}
For pedagogical reasons, in the text we described our procedure estimating selection biases versus spin in the context of a single detector design (initial
LIGO).   In this appendix we provide a similar discussion for advanced and third-generation detectors, specifically the
benchmark advanced LIGO \cite{2010CQGra..27h4006H} and Einstein Telescope \cite{gw-detectors-ET-ScienceDocument} designs.   These instruments' sensitivity is great
enough that, even without detections, their upper limits rule out otherwise astrophysically plausible progenitor
models.  
We particularly emphasize how these detectors superior low-frequency sensitivity leads to much more accurate fits,
over a broader range of masses.
For simplicity, despite both detectors' cosmologically significant range, we perform all calculations in terms of the
\emph{luminosity distance and redshifted mass} --  effectively
as if in a flat universe.  Though our fits  reproduce $\bar{\rho}/\bar{\rho}_o$ to several percent, and though the
reader can invert \emph{any particular line of sight} to that accuracy,  we have not included all information
needed to completely reconstruct the selection bias versus mass and spin.  At these distances,  the detection volume is no longer proportional to a simple cubic moment
of the beampattern function; see for example the appendix of \citet{PSellipticals} for a comparable calculation at low mass.
Though we anticipate the comoving volume swept out by the past detection light cone will not depend sensitively on the
details of its truncation at large redshift\footnote{At large redshift the small amount of comoving volume available at
  a given redshift strongly suppresses lines of sight that reach back to higher redshift.}, we have not thoroughly
explored the errors our neglect of these effects introduces.

By adopting luminosity distance and redshifted mass as parameters, each result in this section is directly comparable to
a corresponding prediction for initial detectors.   However, because advanced detectors have peak sensitivity at roughly $2\times$ lower frequency, the response of initial
detectors of mass $M$ contains comparable waveform content as and is best compared to the response of advanced detectors of mass $2M$.  For example, the low-mass
limit for initial LIGO is roughly $100 M_\odot$ for initial and $200 M_\odot$ for advanced detectors.

\subsection{Average signal power  versus binary spins}
\begin{figure}
\includegraphics{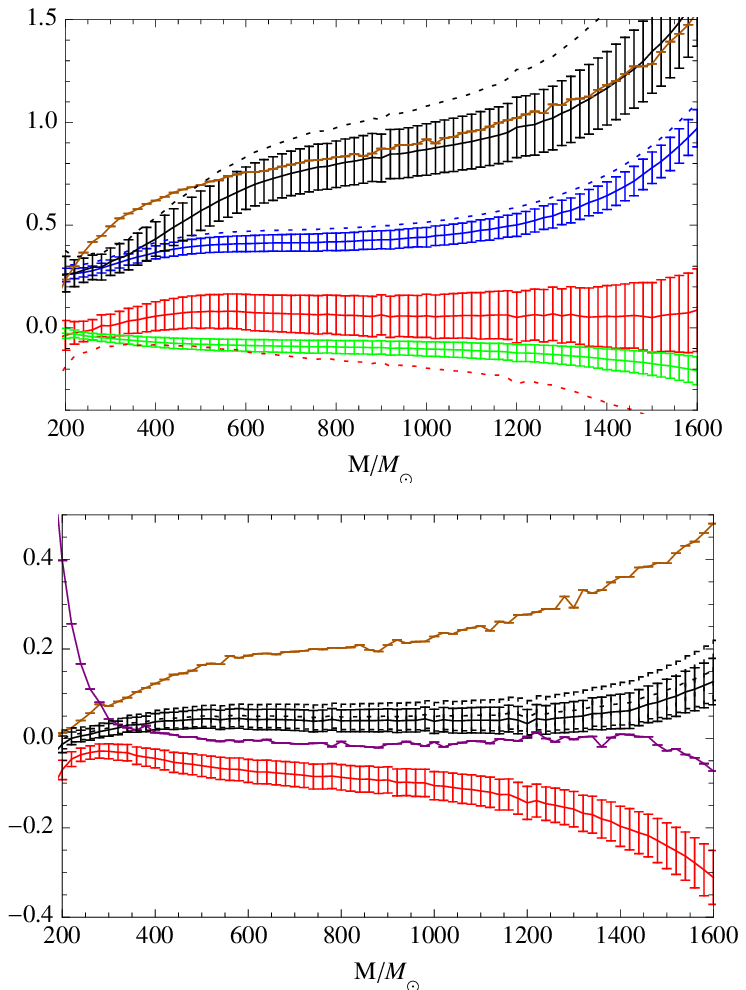}
\caption{\label{fig:Generic:Range:Coefficients:aLIGO} Coefficients of fit to the ratio $\bar{\rho}/\bar{\rho}_o$ for
  single design-sensitivity advanced LIGO detector; compare the top panel to Figures \ref{fig:Generic:FitCoefficients}
  and  \ref{fig:Generic:FitCoefficients}.  The advanced LIGO detector design will be sensitive to lower characteristic
  frequencies [$O(40\unit{Hz})$ vs $O(100\unit{Hz})$ for the initial detector].  For this reason, we provide fits over a
  more limited mass range: only above $200 M_\odot$ will our short
  waveforms start their inspiral well before they enter band. 
}
\end{figure}

\begin{figure}
\includegraphics{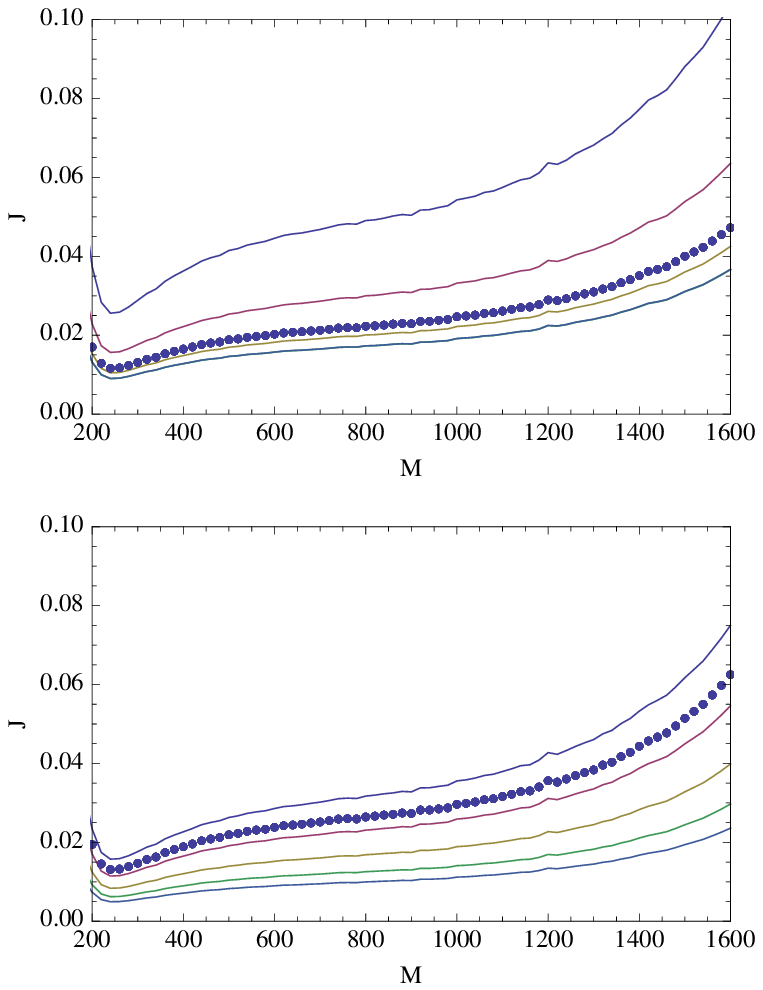}
\caption{\label{fig:Generic:Fit:Error:aLIGO} Expected $1\sigma$ error ${\cal F}$ in an unconstrained
  fit to 
  $F=\bar{\rho}/\bar{\rho}_o$ for advanced LIGO detectors, for spin magnitudes    $|a_1|,|a_2|$ limited to below unity (blue; largest); below $0.8$
  (red); and below  $0.6, 0.4$ and $ 0.2$ (similar
  curves near bottom).  Also shown (points) are the least-squares errors $\delta F$ between our fit $\hat{F}$ and the
  simulated NR data.  Top panel shows
  standard; bottom panel shows error of restricted model.  Because of its broader bandpass, the \emph{advanced}
  detector range is more accurately modeled by our expansion; compare Fig. \ref{fig:ErrorDiagnostics-GlobalVsM-Cubic}.
}
\end{figure}

Unlike the results from initial detectors, the data for $\bar{\rho}/\bar{\rho}_o$ exhibits relatively weak dependence on
spin for all masses.  As a result, a fit to the numerical data performs well,  both reproducing the data 
(Fig. \ref{fig:Generic:Fit:Error:aLIGO})
and producing well-determined fitting coefficients (Fig. \ref{fig:Generic:Range:Coefficients:aLIGO})
 over the entire range of plausible masses
$M\in[200, 1600]M_\odot$.
Aside from a difference in scale, however, the fit exhibits properties comparable to the low-mass fit ($M<500 M_\odot$)
to initial LIGO $\bar{\rho}$.  Notably,
(i) the fit is dominated by aligned spin coefficients, with few resolved corrections involving perpendicular spins;
(ii) it depends only weakly on antisymmetric spin $\chim$;
(iii) it satisfies all sanity conditions, such as increasing monotonically with aligned spin and attaining a local
extrema versus orientation when equal-magnitude spins aligned and
antialigned; 
and finally (iv) higher harmonics $l>3$ only weakly change the best-fit coefficient functions $y_\alpha$, mostly at the highest masses.
Also as in the low-mass fit, the fitting  coefficients are strongly correlated.  Ill-constrained high-order
coefficients like ${\cal B}_{1020}$ are  sensitive to numerical issues, such as extrapolation to large radius.
Most differences between the performance of initial and advanced detectors is directly attributable to their
low-frequency response: advanced detectors  generally lack an abrupt transition from peak to low-frequency
sensitivity.   Without a strong preferred frequency these detectors must have at best a weak bias towards recovering the
largest and most tightly aligned  spins; compare  Section \ref{sec:Results:AlignedSpin}.  Furthermore, as discussed by
example in the Appendix
\ref{ap:AnalyticCoefficients},
because of the nearly power-law low-frequency response of the detector noise power spectrum, the fitting coefficients
are  nearly \emph{constant}, independent of mass.

\begin{figure}
\includegraphics{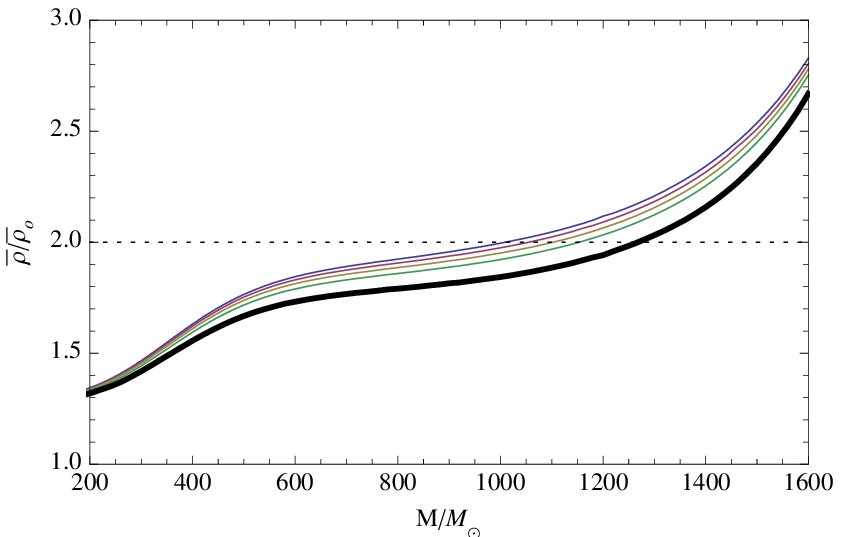}
\caption{\label{fig:AlignedSpin:Diagnostics:aLIGO}
For an equal-mass aligned spin binary with $a_1=a_2=0.8$, estimates for $\bar{\rho}/\bar{\rho}_o$ seen 
by a single advanced LIGO detector at
  $D=100\unit{Mpc}$ versus total binary mass $M$, based on
  waveforms extracted from our numerical merger simulations at 
  coordinate extraction radii
  $r=40,50,60,75$  (blue, red, yellow, green)
and (thick solid) linearly 
  extrapolated in $1/r$ to $r\rightarrow \infty$.
}
\end{figure}

\begin{figure}
\includegraphics{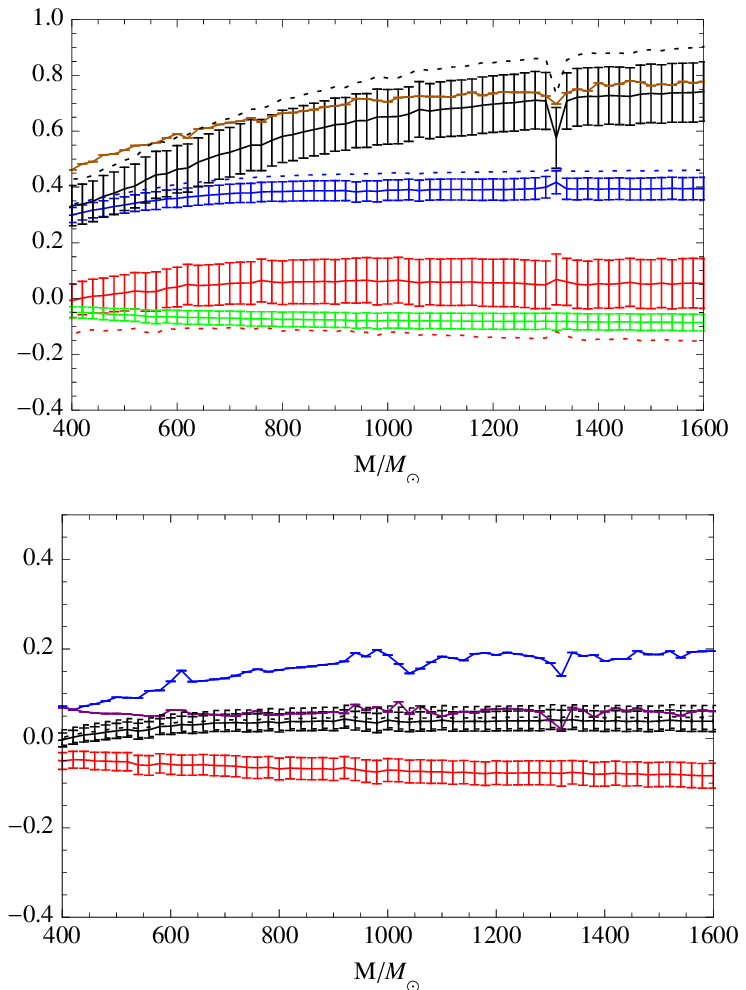}
\caption{\label{fig:Generic:Range:Coefficients:ET} Coefficients of fit to the ratio $\bar{\rho}/\bar{\rho}_o$ for
  single design-sensitivity Einstein Telescope; compare the top panel to Figures \ref{fig:Generic:FitCoefficients}
  and  \ref{fig:Generic:FitCoefficients}.  
}
\end{figure}

\ForInternalReference{
\begin{figure}
\includegraphics[width=\columnwidth]{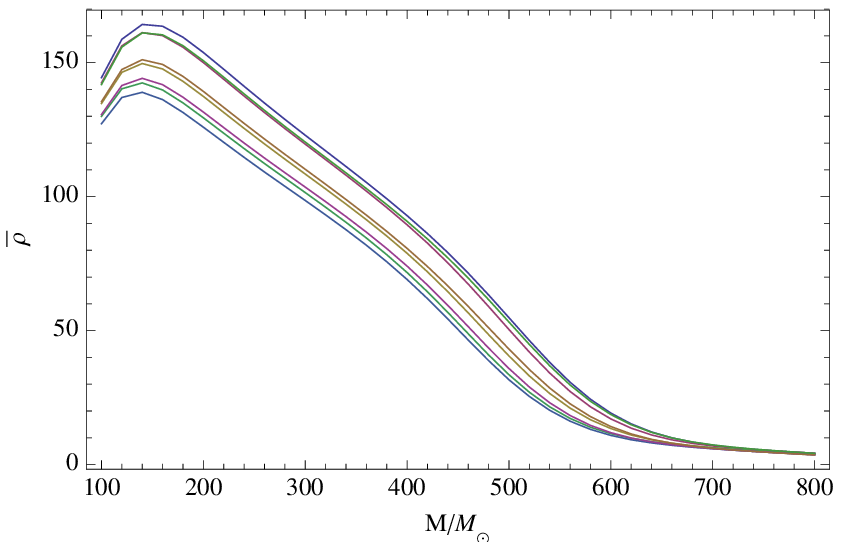}
\caption{\label{fig:SSeries:RangeVersusMass} Angle-averaged SNR $\bar{\rho}$ (top panel) and $\bar{\rho}/\bar{\rho}_o$
  (bottom panel) versus mass for the S-series, with $|a_1|=|a_2|=0.4$.
  Shown (i) to demonstrate my range code is sane, with no lumps or wiggles due to unresolved modes; (ii) for comparison
  with Birjoo papers.
}\end{figure}

\optional{
\subsection{Smoothness and error}
Early drafts of this paper have had aLIGO and ET data show excessive scatter.  These errors are easily traced to large
deviations in individual simulations, where my tabulated $\rho(M)$ is manifestly not smooth at a few data points [Fig \ref{fig:Sanity:Smoothness}].
\begin{figure*}
\includegraphics{fig-mma-SanityCheck-RangesAreSmooth}
\includegraphics{fig-mma-SanityCheck-RangesAreSmooth-aLIGO}
\includegraphics{fig-mma-SanityCheck-RangesAreSmooth-ET}
\caption{\label{fig:Sanity:Smoothness}$\rho(M)$ curves (including $l=4$) for all simulations used, for iLIGO, aLIGO, and ET respectively}
\end{figure*}
\editremark{THESE WILL BE FIXED} -- I presume it's an implementation error/integration failure, perhaps mediated by some
junk radiation causing problems with convergence.

Check the BREAKDOWN PLOTS to see if this can be traced to individual modes

}}

\section{Analytic estimates for expansion coefficients}
\label{ap:AnalyticCoefficients}
As seen in the text, the relative detection volume for different spinning binaries is  largely determined by  the
angle-averaged SNR $\bar{\rho}$.    At both very low and very high mass, the leading-order dependence of the angle-averaged SNR on
spin orientation can be calculated, particularly by adopting high-symmetry configurations such as spin-aligned binaries
to better isolate relevant terms.  For example, at very low mass the aligned spin coefficients such as ${\cal X}_{1,2,3}$ follow from the well-known stationary-phase
Fourier transform of the 2.5-PN-accurate inspiral waveform for spinning, aligned binaries. \optional{ \editremark{citations}.}   Adopting the restricted Newtonian
amplitude at all frequencies and expanding $\sqrt{d\omega/dt}$ in the denominator of the stationary-phase Fourier
transform, the leading order change in SNR with spin follows approximately from
\begin{eqnarray}
 {\cal X}_1 &\approx&  \frac{47}{24} M \pi \frac{J(4/3)}{J(7/3)} \\
{} J(q) &=&  \int_0^\infty \frac{f^{-q}}{S_h}
\end{eqnarray}  
On the other extreme, only the late-stage ringdown of very high-mass spinning binaries will fall into our detectors'
sensitive band.  Qualitatively, ringdown emission produces an SNR limited by the detector frequency  and total amount of
emitted power.  Progenitor black hole spins will change the total amount of emitted energy and, critically, the final
hole ringdown frequency.  Assuming ringdown waves are nearly monochromatic and contain a proportion of the total emitted
energy, the high-mass SNR can be roughly estimated by the nonspinning amplitude at that mass times an  ad-hoc ringdown
correction factor:
\begin{eqnarray}
\bar{\rho} &=&\bar{\rho}_o\frac{\sqrt{(M_f-M)/(M_{f,o}-M)}}{\sqrt{S_h(f_{rd})/S_h(f_{rd,o})}} 
\end{eqnarray}
where $f_{rd}(M_f,a_f)\approx [1-0.63(1-a_f)^{3/10}]/2\pi M_f$ is the ringdown frequency of the final hole of mass $M_f$ and spin $a_f$ and where $M_f,a_f$ are
known expressions of the progenitor masses and spins.  If for simplicity we further retain only   $\chi_{+,z}$ spin
dependence, then we find an expression for ${\cal X}_1$ in terms of the detector's noise power spectrum $S_h$:
\begin{eqnarray}
\bar{\rho}/\bar{\rho}_o %
 &=&1 +\chi_{+,z}\left[0.0888 - 0.0706 \frac{d \ln S_h}{M df}_{f_{rd}}\right] + \ldots \nonumber \\
 &=& 1+ \chi_{+,z}\left[0.0888 + 0.0706 \frac{p \pi (M_f/M)}{(f_{rd,0}M_f)} \right]
\end{eqnarray}
where in the last term we assume the detector's low frequency noise is a pure power law ($S_h(f)\propto f^{-p}$).  The
second term is a positive-definite constant, independent of mass and depending predominantly on the physics of  BH mergers; the detector influences
this estimate for $\bar{\rho}$ only through the low-frequency noise exponent $p$.   
Though not a quantitatively precise model for $\bar{\rho}$ -- the fits described in the paper are required to estimate the detection volume
to astrophysically-relevant accuracy -- this simple ringdown-dominated signal model illustrates why ${\cal
  X}_1$ must vary relatively little  between detector designs.  Our estimate also identifies changes in $d\ln S_h/d\ln f(f=f_{rd}(M))$
as key points where ${\cal X}_1(M)$ should
change significantly.

\section{Detection volume for  networks}
\label{ap:AlternateNetworks}
\label{ap:Beaming}

In the text, we estimated the detection volume to which a single
perpendicular-arm interferometer with a fixed detection threshold could recover a source of unknown sky location.  We
expressed this simple measure of sensitivity  in
terms of two factors ($\bar{\rho}_*,w_*$) .  Real searches employ multiple
detectors, providing   sensitivity to both polarizations nearly everywhere on the sky.  
Their relative orientations determine a sky position- and polarization- dependent sensitivity. Unfortunately, real searches
also have detectors with different (and time-dependent) noise power; a network whose nodes are not always active; 
and of course beampatterns that are highly nonuniform.   A completely realistic treatment of multidetector search
sensitivity adopting realistic search strategies and thresholds is beyond the scope of this paper.

Though necessarily incomplete, the analytic estimates of idealized detector performance nonetheless allow us to
quickly understand the dominant qualitative effects spin might have on the detection volume, such as Fig. \ref{fig:astro:DetectionVolumeVersusSpin}.
Similar analytic estimates for idealized \emph{networks} can be constructed, albeit depending on network topology.    For
coherent multidetector searches with identical detectors, their sensitivity along any direction is
described in general by a 2x2 Hermetian matrix for each sky position  \cite{CutlerFlanagan:1994}; for brevity, we will
limit attention to this case henceforth.     This orientation- and polarization-dependent sensitivity
implies the \emph{true} beampattern function $w$ of both source- and sky position- orientations becomes very complex in
general, particularly when multiple harmonics with different shapes contribute to the overall response.    Nonetheless,
there are three physically interesting approximations where the average amplitude and beampattern are tractable:
\begin{itemize}
\item \emph{Single detector} (used in text):   Sensitivity is characterized by an angle-averaged SNR $\bar{\rho}_*$ [Eq. (\ref{eq:def:rhobar:star})]; a
  beampattern function $w_* = \rho/\bar{\rho}_*$ of sky position and source orientation; and a beampattern
  effective radius $\bar{w}_*=\left<w_*^3\right>^{1/3}$ [Eq. (\ref{eq:def:wbar:star})],
  averaged over all orientations.   

To provide a concrete example, consider the case of a binary dominated by $l=|m|=2$ emission, viewed by a single
interferometer.  Since the beampattern $w$ versus all angles  is known analytically [e.g., Eq. (2) in
\cite{PSellipticals}], the amplitude $\rho$ versus orientation is completely determined by the optimally-oriented
amplitude $\rho_{\rm *, max}$, or equivalently by the ``horizon distance'' $D_{\rm h}$ at which $\rho_{max}=\rho_c$.
\begin{eqnarray}
\bar{\rho}_* &=&  \rho_{*,max} \frac{2}{5} =  \rho_c\frac{ 2}{5}\frac{D_h}{D} \\
\bar{D}_*    &=& D_h 2/5 \\
\bar{w}_*    &=&  \wStarNonspinning 
\end{eqnarray}
Alternatively, the sensitivity of the interferometer can be characterized by source-orientation-averaged effective detection
volume, or,  to use similar units, the \emph{volume-averaged distance}  $D_{\rm v}$.  The above expressions allow us to
express familiar relationship between $D_{\rm v}$ and $D_{\rm h}$ \cite{PSellipticals} in terms of the single-detector
beampattern correction factor $\bar{w}_*$:
\begin{eqnarray}
D_v &=& D_h 2 \bar{w}_*/ 5 \approx D_h/(2.26)
\end{eqnarray}

\item \emph{Isotropic network, equal polarization sensitivity }: At the other extreme from a single-detector network is
  the ideal case: a network with equal sensitivity to both polarizations in all directions.   Assuming searches for this
  ideal network are carried out by matched filtering, its network sensitivity is characterized by  an angle-averaged
  two-polarization network SNR $\bar{\rho}= \bar{\rho}_* \sqrt{5}$
  [Eq. (\ref{eq:def:rhobar}); see \citet{CutlerFlanagan:1994}]; a
  beampattern function $w=\rho/\bar{\rho}$; and a \emph{source-orientation-averaged} moment
\begin{eqnarray}
\label{eq:def:wbar}
\bar{w} =  &=&   \frac{[\int \frac{d\Omega_n}{4 \pi} w(\myvector{\hat{n}})^3]^{1/3}}{\left<w^2\right>^{2/3} } \; .
\end{eqnarray}
where, because the detector has equal sensitivity to both polarizations, the orientation average needs to be conducted
only over all emission directions $\myvector{\hat{n}}$.

\item \emph{Isotropic network, single polarization sensitivity}:  In between these two extremes is an isotropic network
  with sensitivity to only \emph{one} polarization at each sky position.  For example, the three-site LIGO-Virgo network is primarily sensitive to one
  polarization for each sky position.   Using a subscript $0$ to denote this class of detector, the the beampattern function
  can be similarly defined as a suitable average over emission direction $\myvector{\hat{n}}$ and polarization angle $\psi$:
\begin{eqnarray}
\label{eq:def:wbar:zero}
\bar{w}_0 &=& \left<w_0^3\right>^{1/3}/\left<w_0^2\right>
\end{eqnarray}
By symmetry, the orientation-averaged signal power  must be precisely half that of a network sensitive to two
polarizations, so $\bar{\rho}_0=\bar{\rho}/\sqrt{2}$.

\end{itemize}

To illustrate the beaming-induced differences in detection volume  between these different network topologies,  in Table \ref{tab:CanonicalBeamingFactors} we tabulate the beaming correction factors when a particular
multipolar mode dominates emission ($l=|m|$).   All are nearly unity; all depend weakly on $l$, for a given detector
topology.     Additionally, the beaming correction factors have a universal hierarchy: $\bar{w}_o>\bar{w}_*>\bar{w}$.
Obviously the idealized two-polarization-sensitive isotropic network has the most symmetric beampattern, least sensitive
to host orientation and polarization content.  Conversely, an isotropic network sensitive to \emph{one} polarization is
usually 
the \emph{most} sensitive to unfortunately oriented (polarized) sources.  

\begin{table}
\input{tab-mma-CanonicalBeamingFactors.tex}
\caption{\label{tab:CanonicalBeamingFactors}  Beaming correction factors when a single $l=|m|$ mode dominates emission.  
In this table, $\bar{w}_*$ times $1/\sqrt{5}$ characterizes the radius of the detection volume associated with an all-sky search for a single source by a
single interferometer; $\bar{w}$ (without additional factors) refers to a comparable search by a network with isotropic sensitivity to both
polarizations; and $\bar{w}_o$ refers to a network with isotropic sensitivity to a single polarization.
}
\end{table}

\optional{ADDED ORGANIZATION

* key points

 -- transition idea from one regime to another

\subsection{Beaming at fixed time, frequency}

* import discussion from notes

\subsection{Beaming and mode content}

}

\ForInternalReference{
\section{Implementation details: NR data to $\bar{\rho},w$}

* the stuff Birjoo worries about: 

 - truncation in amplitude at late times.  

  - fourier padding.

 - Starting frequency for higher modes: a possible slight bias introduced against them

}

\bibliography{bibexport,nr,ligo}
\end{document}

%% file: tab-mma-MomentMatrix.tex
\begin{tabular}{llllllllll}
 
 & ${\cal X}_1$ & ${\cal X}_2$ & ${\cal X}_3$ & ${\cal X}_{02}$ & ${\cal A}_{20}$ & ${\cal A}_{02}$ & ${\cal B}_{1200}$ & ${\cal B}_{1020}$ & ${\cal B}_{1002}$\\  & 0.168 & 0. & 0.065 & 0. & 0. & 0. & 0.047 & 0. & 0.047  \\ 
${\cal X}_1$ & 0. & 0.065 & 0. & 0.047 & 0. & 0.047 & 0. & 0. & 0.  \\ 
${\cal X}_2$ & 0.065 & 0. & 0.032 & 0. & 0. & 0. & 0.015 & 0. & 0.015  \\ 
${\cal X}_3$ & 0. & 0.047 & 0. & 0.175 & 0.046 & 0.064 & 0. & 0. & 0.  \\ 
${\cal X}_{02}$ & 0. & 0. & 0. & 0.046 & 0.063 & 0.047 & 0. & 0. & 0.  \\ 
${\cal A}_{20}$ & 0. & 0.047 & 0. & 0.064 & 0.047 & 0.176 & 0. & 0. & 0.  \\ 
${\cal A}_{02}$ & 0.047 & 0. & 0.015 & 0. & 0. & 0. & 0.021 & 0. & 0.  \\ 
${\cal B}_{1200}$ & 0. & 0. & 0. & 0. & 0. & 0. & 0. & 0. & 0.  \\ 
${\cal B}_{1020}$ & 0.047 & 0. & 0.015 & 0. & 0. & 0. & 0. & 0. & 0.021  
 \end{tabular}

%% file: tab-mma-SimulationSet.tex
\begin{tabular}{llllllllll}
$S_{1,x}/M^2$ & $S_{1,y}/M^2$ & $S_{1,z}/M^2$ & $S_{2,x}/M^2$ & $S_{2,y}/M^2$ & $S_{2,z}/M^2$ & $T/M$ & $T_{wave}/M$ & $r_{start}$ & $M/h_{min}$\\ 
 -0.0785 & 0.044 & 0.12 & 0.1 & -0.11 & -0.02 & 1297.2 & 872.3 & 10. & 77 \\
 0.0381 & -0.0565 & 0.1337 & -0.112 & -0.0258 & -0.0964 & 1493.8 & 784.9 & 10. & 77 \\
 0. & 0. & 0.05 & 0. & 0. & -0.05 & 1398.9 & 742.6 & 10. & 77 \\
 0. & 0. & 0.1 & 0. & 0. & -0.1 & 1398.9 & 732.4 & 10. & 77 \\
 0. & 0. & 0.15 & 0. & 0. & -0.15 & 1398.9 & 720.9 & 10. & 77 \\
 0. & 0. & 0.2 & 0. & 0. & -0.2 & 1398.9 & 687. & 10. & 77 \\
 0. & 0. & 0.2 & 0. & 0. & 0.2 & 1499. & 1389.5 & 10. & 77 \\
 0. & 0. & 0. & 0. & 0. & 0.2 & 1510.3 & 1201.7 & 10. & 77 \\
 0. & 0. & 0. & 0. & 0. & 0. & 1099. & 885.5 & 10. & 77 \\
 0. & 0. & 0.05 & 0. & 0. & 0.05 & 1298.9 & 906.4 & 10. & 77 \\
 0. & 0. & 0.05 & 0.0354 & 0. & 0.0354 & 1298.9 & 849. & 10. & 77 \\
 0. & 0. & 0.05 & 0.0433 & 0. & 0.025 & 1298.9 & 837. & 10. & 77 \\
 0. & 0. & 0.05 & 0.05 & 0. & 0. & 1298.9 & 838.2 & 10. & 77 \\
 0. & 0. & 0.1 & 0. & 0. & 0.1 & 1398.9 & 997.5 & 10. & 77 \\
 0. & 0. & 0.1 & 0.0707 & 0. & 0.0707 & 1499. & 986.6 & 10. & 77 \\
 0. & 0. & 0.1 & 0.0866 & 0. & 0.05 & 1499. & 965.8 & 10. & 77 \\
 0. & 0. & 0.1 & 0.1 & 0. & 0. & 1499. & 908.3 & 10. & 77 \\
 0. & 0. & 0.15 & 0. & 0. & 0.15 & 1499. & 1086. & 10. & 77 \\
 0. & 0. & 0.15 & 0.1061 & 0. & 0.1061 & 1499. & 1052.1 & 10. & 77 \\
 0. & 0. & 0.15 & 0.1299 & 0. & 0.075 & 1499. & 993.7 & 10. & 77 \\
 0. & 0. & 0.15 & 0.15 & 0. & 0. & 1499. & 915.7 & 10. & 77 \\
 0. & 0. & 0.2 & 0. & 0. & 0.2 & 1699. & 1217.7 & 10. & 77 \\
 0. & 0. & 0.05 & -0.05 & 0. & 0. & 392.7 & 250.5 & 6.2 & 77 \\
 0.0354 & 0. & 0.0354 & -0.05 & 0. & 0. & 380.7 & 246.6 & 6.2 & 77 \\
 0.05 & 0. & 0. & -0.05 & 0. & 0. & 389.2 & 217.4 & 6.2 & 77 \\
 0.0354 & 0. & -0.0354 & -0.05 & 0. & 0. & 394.9 & 206.9 & 6.2 & 77 \\
 0. & 0. & -0.05 & -0.05 & 0. & 0. & 394.9 & 210.8 & 6.2 & 77 \\
 -0.0354 & 0. & -0.0354 & -0.05 & 0. & 0. & 389.2 & 207.9 & 6.2 & 77 \\
 -0.05 & 0. & 0. & -0.05 & 0. & 0. & 383.2 & 221.8 & 6.2 & 77 \\
 -0.0354 & 0. & 0.0354 & -0.05 & 0. & 0. & 384. & 247.6 & 6.2 & 77 \\
 0. & 0. & 0.1 & -0.1 & 0. & 0. & 398.9 & 277.3 & 6.2 & 77 \\
 0.0707 & 0. & 0.0707 & -0.1 & 0. & 0. & 398.9 & 264.5 & 6.2 & 77 \\
 0.1 & 0. & 0. & -0.1 & 0. & 0. & 398.9 & 227.1 & 6.2 & 77 \\
 0.0707 & 0. & -0.0707 & -0.1 & 0. & 0. & 398.9 & 209.8 & 6.2 & 77 \\
 0. & 0. & -0.1 & -0.1 & 0. & 0. & 398.9 & 208.7 & 6.2 & 77 \\
 -0.0707 & 0. & -0.0707 & -0.1 & 0. & 0. & 398.9 & 209.6 & 6.2 & 77 \\
 -0.1 & 0. & 0. & -0.1 & 0. & 0. & 398.9 & 240.8 & 6.2 & 77 \\
 -0.0707 & 0. & 0.0707 & -0.1 & 0. & 0. & 398.9 & 274.2 & 6.2 & 77 \\
 0. & 0. & 0.15 & -0.15 & 0. & 0. & 398.9 & 289.7 & 6.2 & 77 \\
 0.0388 & 0. & 0.1449 & -0.15 & 0. & 0. & 498.9 & 288.7 & 6.2 & 77 \\
 0.075 & 0. & 0.1299 & -0.15 & 0. & 0. & 498.9 & 276.7 & 6.2 & 77 \\
 0.1061 & 0. & 0.1061 & -0.15 & 0. & 0. & 498.9 & 270.5 & 6.2 & 77 \\
 0.1299 & 0. & 0.075 & -0.15 & 0. & 0. & 498.9 & 263.3 & 6.2 & 77 \\
 0.1449 & 0. & 0.0388 & -0.15 & 0. & 0. & 498.9 & 248.2 & 6.2 & 77 \\
 0.15 & 0. & 0. & -0.15 & 0. & 0. & 398.9 & 229.8 & 6.2 & 77 \\
 0.1449 & 0. & -0.0388 & -0.15 & 0. & 0. & 498.9 & 200.5 & 6.2 & 77 \\
 0.1299 & 0. & -0.075 & -0.15 & 0. & 0. & 498.9 & 211. & 6.2 & 77 \\
 0.1061 & 0. & -0.1061 & -0.15 & 0. & 0. & 498.9 & 206.9 & 6.2 & 77 \\
 0.075 & 0. & -0.1299 & -0.15 & 0. & 0. & 498.9 & 202.1 & 6.2 & 77 \\
 0.0388 & 0. & -0.1449 & -0.15 & 0. & 0. & 498.9 & 198.6 & 6.2 & 77 \\
 0. & 0. & -0.15 & -0.15 & 0. & 0. & 498.9 & 198.4 & 6.2 & 77 \\
 -0.0388 & 0. & -0.1449 & -0.15 & 0. & 0. & 498.9 & 199.8 & 6.2 & 77 \\
 -0.075 & 0. & -0.1299 & -0.15 & 0. & 0. & 498.9 & 204.6 & 6.2 & 77 \\
 -0.1061 & 0. & -0.1061 & -0.15 & 0. & 0. & 498.9 & 206.5 & 6.2 & 77 \\
 -0.1299 & 0. & -0.075 & -0.15 & 0. & 0. & 498.9 & 204. & 6.2 & 77 \\
 -0.1449 & 0. & -0.0388 & -0.15 & 0. & 0. & 495.6 & 233.1 & 6.2 & 77 \\
 -0.15 & 0. & 0. & -0.15 & 0. & 0. & 436.7 & 254.4 & 6.2 & 77 \\
 -0.1449 & 0. & 0.0388 & -0.15 & 0. & 0. & 488.4 & 274.2 & 6.2 & 77 \\
 -0.1299 & 0. & 0.075 & -0.15 & 0. & 0. & 498.9 & 279.2 & 6.2 & 77 \\
 -0.1061 & 0. & 0.1061 & -0.15 & 0. & 0. & 498.9 & 287.7 & 6.2 & 77 \\
 -0.075 & 0. & 0.1299 & -0.15 & 0. & 0. & 498.9 & 289.5 & 6.2 & 77 \\
 -0.0388 & 0. & 0.1449 & -0.15 & 0. & 0. & 498.9 & 290.2 & 6.2 & 77 \\
 -0.1477 & 0. & -0.026 & -0.15 & 0. & 0. & 498.9 & 248. & 6.2 & 77 \\
 -0.1494 & 0. & -0.0131 & -0.15 & 0. & 0. & 498.9 & 251.3 & 6.2 & 77
\end{tabular}

%% file: tab-mma-CanonicalBeamingFactors.tex
\begin{tabular}{llll}
l & $\bar{w}_o$ & $\bar{w}_*$ & $\bar{w}$\\ 
 2 & 1.11661 & 1.09664 & 1.05065 \\
 4 & 1.10614 & 1.08541 & 1.0408 \\
 6 & 1.15301 & 1.12961 & 1.08489
\end{tabular}

%% file: paper.bbl
\begin{thebibliography}{54}
\expandafter\ifx\csname natexlab\endcsname\relax\def\natexlab#1{#1}\fi
\expandafter\ifx\csname bibnamefont\endcsname\relax
  \def\bibnamefont#1{#1}\fi
\expandafter\ifx\csname bibfnamefont\endcsname\relax
  \def\bibfnamefont#1{#1}\fi
\expandafter\ifx\csname citenamefont\endcsname\relax
  \def\citenamefont#1{#1}\fi
\expandafter\ifx\csname url\endcsname\relax
  \def\url#1{\texttt{#1}}\fi
\expandafter\ifx\csname urlprefix\endcsname\relax\def\urlprefix{URL }\fi
\providecommand{\bibinfo}[2]{#2}
\providecommand{\eprint}[2][]{\url{#2}}

\bibitem[{\citenamefont{{Acernese} et~al.}(2006)\citenamefont{{Acernese},
  {Amico}, {Alshourbagy}, {Antonucci}, {Aoudia}, {Avino}, {Babusci},
  {Ballardin}, {Barone}, {Barsotti} et~al.}}]{gw-detectors-VIRGO-original}
\bibinfo{author}{\bibfnamefont{F.}~\bibnamefont{{Acernese}}},
  \bibinfo{author}{\bibfnamefont{P.}~\bibnamefont{{Amico}}},
  \bibinfo{author}{\bibfnamefont{M.}~\bibnamefont{{Alshourbagy}}},
  \bibinfo{author}{\bibfnamefont{F.}~\bibnamefont{{Antonucci}}},
  \bibinfo{author}{\bibfnamefont{S.}~\bibnamefont{{Aoudia}}},
  \bibinfo{author}{\bibfnamefont{S.}~\bibnamefont{{Avino}}},
  \bibinfo{author}{\bibfnamefont{D.}~\bibnamefont{{Babusci}}},
  \bibinfo{author}{\bibfnamefont{G.}~\bibnamefont{{Ballardin}}},
  \bibinfo{author}{\bibfnamefont{F.}~\bibnamefont{{Barone}}},
  \bibinfo{author}{\bibfnamefont{L.}~\bibnamefont{{Barsotti}}},
  \bibnamefont{et~al.}, \bibinfo{journal}{Classical and Quantum Gravity}
  \textbf{\bibinfo{volume}{23}}, \bibinfo{pages}{635} (\bibinfo{year}{2006}).

\bibitem[{\citenamefont{{Abbott et al. (The LIGO Scientific
  Collaboration)}}(2003)}]{gw-detectors-LIGO-original}
\bibinfo{author}{\bibnamefont{{Abbott et al. (The LIGO Scientific
  Collaboration)}}}, \bibinfo{journal}{(gr-qc/0308043)}
  (\bibinfo{year}{2003}),
  \urlprefix\url{http://xxx.lanl.gov/abs/gr-qc/0308043}.

\bibitem[{\citenamefont{{Shoemaker} and {the LIGO Scientific
  Collaboration}}(2009)}]{LIGO-aLIGODesign-Sensitivity}
\bibinfo{author}{\bibfnamefont{D.}~\bibnamefont{{Shoemaker}}} \bibnamefont{and}
  \bibinfo{author}{\bibnamefont{{the LIGO Scientific Collaboration}}}
  (\bibinfo{year}{2009}),
  \urlprefix\url{https://dcc.ligo.org/cgi-bin/DocDB/ShowDocument?docid=2974}.

\bibitem[{\citenamefont{{O'Shaughnessy}
  et~al.}(2009)\citenamefont{{O'Shaughnessy}, {Kalogera}, and
  {Belcynski}}}]{PSellipticals}
\bibinfo{author}{\bibfnamefont{R.}~\bibnamefont{{O'Shaughnessy}}},
  \bibinfo{author}{\bibfnamefont{V.}~\bibnamefont{{Kalogera}}},
  \bibnamefont{and}
  \bibinfo{author}{\bibfnamefont{K.}~\bibnamefont{{Belcynski}}},
  \bibinfo{journal}{(arXiv:0908.3635) Submitted to ApJ}
  (\bibinfo{year}{2009}), \eprint{0908.3635}.

\bibitem[{\citenamefont{{Belczynski} et~al.}(2010)\citenamefont{{Belczynski},
  {Dominik}, {Bulik}, {O'Shaughnessy}, {Fryer}, and
  {Holz}}}]{popsyn-LowMetallicityImpact-Chris2010}
\bibinfo{author}{\bibfnamefont{K.}~\bibnamefont{{Belczynski}}},
  \bibinfo{author}{\bibfnamefont{M.}~\bibnamefont{{Dominik}}},
  \bibinfo{author}{\bibfnamefont{T.}~\bibnamefont{{Bulik}}},
  \bibinfo{author}{\bibfnamefont{R.}~\bibnamefont{{O'Shaughnessy}}},
  \bibinfo{author}{\bibfnamefont{C.~L.} \bibnamefont{{Fryer}}},
  \bibnamefont{and} \bibinfo{author}{\bibfnamefont{D.~E.}
  \bibnamefont{{Holz}}}, \bibinfo{journal}{(arXiv:1004.0386)}
  (\bibinfo{year}{2010}), \eprint{1004.0386}.

\bibitem[{\citenamefont{{O'Shaughnessy}
  et~al.}(2007)\citenamefont{{O'Shaughnessy}, {O'Leary}, and
  {Rasio}}}]{clusters-2005}
\bibinfo{author}{\bibfnamefont{R.}~\bibnamefont{{O'Shaughnessy}}},
  \bibinfo{author}{\bibfnamefont{R.}~\bibnamefont{{O'Leary}}},
  \bibnamefont{and} \bibinfo{author}{\bibfnamefont{F.~A.}
  \bibnamefont{{Rasio}}}, \bibinfo{journal}{\prd}
  \textbf{\bibinfo{volume}{76}}, \bibinfo{pages}{061504}
  (\bibinfo{year}{2007}), \eprint{0701887}.

\bibitem[{\citenamefont{{Sadowski} et~al.}(2008)\citenamefont{{Sadowski},
  {Belczynski}, {Bulik}, {Ivanova}, {Rasio}, and
  {O'Shaughnessy}}}]{2008ApJ...676.1162S}
\bibinfo{author}{\bibfnamefont{A.}~\bibnamefont{{Sadowski}}},
  \bibinfo{author}{\bibfnamefont{K.}~\bibnamefont{{Belczynski}}},
  \bibinfo{author}{\bibfnamefont{T.}~\bibnamefont{{Bulik}}},
  \bibinfo{author}{\bibfnamefont{N.}~\bibnamefont{{Ivanova}}},
  \bibinfo{author}{\bibfnamefont{F.~A.} \bibnamefont{{Rasio}}},
  \bibnamefont{and}
  \bibinfo{author}{\bibfnamefont{R.}~\bibnamefont{{O'Shaughnessy}}},
  \bibinfo{journal}{\apj} \textbf{\bibinfo{volume}{676}}, \bibinfo{pages}{1162}
  (\bibinfo{year}{2008}).

\bibitem[{\citenamefont{{Banerjee} et~al.}(2010)\citenamefont{{Banerjee},
  {Baumgardt}, and {Kroupa}}}]{2010MNRAS.402..371B}
\bibinfo{author}{\bibfnamefont{S.}~\bibnamefont{{Banerjee}}},
  \bibinfo{author}{\bibfnamefont{H.}~\bibnamefont{{Baumgardt}}},
  \bibnamefont{and} \bibinfo{author}{\bibfnamefont{P.}~\bibnamefont{{Kroupa}}},
  \bibinfo{journal}{\mnras} \textbf{\bibinfo{volume}{402}},
  \bibinfo{pages}{371} (\bibinfo{year}{2010}), \eprint{0910.3954}.

\bibitem[{\citenamefont{{Fregeau} et~al.}(2006)\citenamefont{{Fregeau},
  {Larson}, {Miller}, {O'Shaughnessy}, and {Rasio}}}]{imbhlisa-2006}
\bibinfo{author}{\bibfnamefont{J.~M.} \bibnamefont{{Fregeau}}},
  \bibinfo{author}{\bibfnamefont{S.~L.} \bibnamefont{{Larson}}},
  \bibinfo{author}{\bibfnamefont{M.~C.} \bibnamefont{{Miller}}},
  \bibinfo{author}{\bibfnamefont{R.}~\bibnamefont{{O'Shaughnessy}}},
  \bibnamefont{and} \bibinfo{author}{\bibfnamefont{F.~A.}
  \bibnamefont{{Rasio}}}, \bibinfo{journal}{\apj}
  \textbf{\bibinfo{volume}{646}}, \bibinfo{pages}{L135} (\bibinfo{year}{2006}),
  \urlprefix\url{http://xxx.lanl.gov/abs/astro-ph/0605732}.

\bibitem[{\citenamefont{{Lousto} et~al.}(2010)\citenamefont{{Lousto}, {Nakano},
  {Zlochower}, and {Campanelli}}}]{2010PhRvD..81h4023L}
\bibinfo{author}{\bibfnamefont{C.~O.} \bibnamefont{{Lousto}}},
  \bibinfo{author}{\bibfnamefont{H.}~\bibnamefont{{Nakano}}},
  \bibinfo{author}{\bibfnamefont{Y.}~\bibnamefont{{Zlochower}}},
  \bibnamefont{and}
  \bibinfo{author}{\bibfnamefont{M.}~\bibnamefont{{Campanelli}}},
  \bibinfo{journal}{\prd} \textbf{\bibinfo{volume}{81}},
  \bibinfo{pages}{084023} (\bibinfo{year}{2010}), \eprint{0910.3197}.

\bibitem[{\citenamefont{{Campanelli} et~al.}(2009)\citenamefont{{Campanelli},
  {Lousto}, {Nakano}, and {Zlochower}}}]{2009PhRvD..79h4010C}
\bibinfo{author}{\bibfnamefont{M.}~\bibnamefont{{Campanelli}}},
  \bibinfo{author}{\bibfnamefont{C.~O.} \bibnamefont{{Lousto}}},
  \bibinfo{author}{\bibfnamefont{H.}~\bibnamefont{{Nakano}}}, \bibnamefont{and}
  \bibinfo{author}{\bibfnamefont{Y.}~\bibnamefont{{Zlochower}}},
  \bibinfo{journal}{\prd} \textbf{\bibinfo{volume}{79}},
  \bibinfo{pages}{084010} (\bibinfo{year}{2009}), \eprint{0808.0713}.

\bibitem[{\citenamefont{{Szil{\'a}gyi}
  et~al.}(2009)\citenamefont{{Szil{\'a}gyi}, {Lindblom}, and
  {Scheel}}}]{2009PhRvD..80l4010S}
\bibinfo{author}{\bibfnamefont{B.}~\bibnamefont{{Szil{\'a}gyi}}},
  \bibinfo{author}{\bibfnamefont{L.}~\bibnamefont{{Lindblom}}},
  \bibnamefont{and} \bibinfo{author}{\bibfnamefont{M.~A.}
  \bibnamefont{{Scheel}}}, \bibinfo{journal}{\prd}
  \textbf{\bibinfo{volume}{80}}, \bibinfo{pages}{124010}
  (\bibinfo{year}{2009}), \eprint{0909.3557}.

\bibitem[{\citenamefont{{Campanelli} et~al.}(2010)\citenamefont{{Campanelli},
  {Lousto}, {Mundim}, {Nakano}, {Zlochower}, and
  {Bischof}}}]{2010CQGra..27h4034C}
\bibinfo{author}{\bibfnamefont{M.}~\bibnamefont{{Campanelli}}},
  \bibinfo{author}{\bibfnamefont{C.~O.} \bibnamefont{{Lousto}}},
  \bibinfo{author}{\bibfnamefont{B.~C.} \bibnamefont{{Mundim}}},
  \bibinfo{author}{\bibfnamefont{H.}~\bibnamefont{{Nakano}}},
  \bibinfo{author}{\bibfnamefont{Y.}~\bibnamefont{{Zlochower}}},
  \bibnamefont{and}
  \bibinfo{author}{\bibfnamefont{H.}~\bibnamefont{{Bischof}}},
  \bibinfo{journal}{Classical and Quantum Gravity}
  \textbf{\bibinfo{volume}{27}}, \bibinfo{pages}{084034}
  (\bibinfo{year}{2010}), \eprint{1001.3834}.

\bibitem[{\citenamefont{Rezzolla}(2009)}]{gr-nr-io-review-Rezzolla2008}
\bibinfo{author}{\bibfnamefont{L.}~\bibnamefont{Rezzolla}},
  \bibinfo{journal}{Classical and Quantum Gravity}
  \textbf{\bibinfo{volume}{26}}, \bibinfo{pages}{094023}
  (\bibinfo{year}{2009}), \eprint{0812.2325}.

\bibitem[{\citenamefont{{Barausse} and
  {Rezzolla}}(2009)}]{gr-nr-io-fitting-BR2009}
\bibinfo{author}{\bibfnamefont{E.}~\bibnamefont{{Barausse}}} \bibnamefont{and}
  \bibinfo{author}{\bibfnamefont{L.}~\bibnamefont{{Rezzolla}}},
  \bibinfo{journal}{\apjl} \textbf{\bibinfo{volume}{704}}, \bibinfo{pages}{L40}
  (\bibinfo{year}{2009}), \eprint{0904.2577}.

\bibitem[{\citenamefont{{Marronetti} et~al.}(2008)\citenamefont{{Marronetti},
  {Tichy}, {Br{\"u}gmann}, {Gonz{\'a}lez}, and
  {Sperhake}}}]{2008PhRvD..77f4010M}
\bibinfo{author}{\bibfnamefont{P.}~\bibnamefont{{Marronetti}}},
  \bibinfo{author}{\bibfnamefont{W.}~\bibnamefont{{Tichy}}},
  \bibinfo{author}{\bibfnamefont{B.}~\bibnamefont{{Br{\"u}gmann}}},
  \bibinfo{author}{\bibfnamefont{J.}~\bibnamefont{{Gonz{\'a}lez}}},
  \bibnamefont{and}
  \bibinfo{author}{\bibfnamefont{U.}~\bibnamefont{{Sperhake}}},
  \bibinfo{journal}{\prd} \textbf{\bibinfo{volume}{77}},
  \bibinfo{pages}{064010} (\bibinfo{year}{2008}), \eprint{0709.2160}.

\bibitem[{\citenamefont{{Kesden} et~al.}(2010)\citenamefont{{Kesden},
  {Sperhake}, and {Berti}}}]{2010PhRvD..81h4054K}
\bibinfo{author}{\bibfnamefont{M.}~\bibnamefont{{Kesden}}},
  \bibinfo{author}{\bibfnamefont{U.}~\bibnamefont{{Sperhake}}},
  \bibnamefont{and} \bibinfo{author}{\bibfnamefont{E.}~\bibnamefont{{Berti}}},
  \bibinfo{journal}{\prd} \textbf{\bibinfo{volume}{81}},
  \bibinfo{pages}{084054} (\bibinfo{year}{2010}), \eprint{1002.2643}.

\bibitem[{\citenamefont{{Ajith} et~al.}(2009)\citenamefont{{Ajith}, {Hannam},
  {Husa}, {Chen}, {Bruegmann}, {Dorband}, {Mueller}, {Ohme}, {Pollney},
  {Reisswig} et~al.}}]{gwastro-Ajith-AlignedSpinWaveforms}
\bibinfo{author}{\bibfnamefont{P.}~\bibnamefont{{Ajith}}},
  \bibinfo{author}{\bibfnamefont{M.}~\bibnamefont{{Hannam}}},
  \bibinfo{author}{\bibfnamefont{S.}~\bibnamefont{{Husa}}},
  \bibinfo{author}{\bibfnamefont{Y.}~\bibnamefont{{Chen}}},
  \bibinfo{author}{\bibfnamefont{B.}~\bibnamefont{{Bruegmann}}},
  \bibinfo{author}{\bibfnamefont{N.}~\bibnamefont{{Dorband}}},
  \bibinfo{author}{\bibfnamefont{D.}~\bibnamefont{{Mueller}}},
  \bibinfo{author}{\bibfnamefont{F.}~\bibnamefont{{Ohme}}},
  \bibinfo{author}{\bibfnamefont{D.}~\bibnamefont{{Pollney}}},
  \bibinfo{author}{\bibfnamefont{C.}~\bibnamefont{{Reisswig}}},
  \bibnamefont{et~al.}, \bibinfo{journal}{(arXiv:0909.2867)}
  (\bibinfo{year}{2009}), \eprint{0909.2867}.

\bibitem[{\citenamefont{{Reisswig}
  et~al.}(2009{\natexlab{a}})\citenamefont{{Reisswig}, {Husa}, {Rezzolla},
  {Dorband}, {Pollney}, and {Seiler}}}]{gwastro-nr-AlignedSpinVolumeWeight}
\bibinfo{author}{\bibfnamefont{C.}~\bibnamefont{{Reisswig}}},
  \bibinfo{author}{\bibfnamefont{S.}~\bibnamefont{{Husa}}},
  \bibinfo{author}{\bibfnamefont{L.}~\bibnamefont{{Rezzolla}}},
  \bibinfo{author}{\bibfnamefont{E.~N.} \bibnamefont{{Dorband}}},
  \bibinfo{author}{\bibfnamefont{D.}~\bibnamefont{{Pollney}}},
  \bibnamefont{and} \bibinfo{author}{\bibfnamefont{J.}~\bibnamefont{{Seiler}}},
  \bibinfo{journal}{\prd} \textbf{\bibinfo{volume}{80}},
  \bibinfo{pages}{124026} (\bibinfo{year}{2009}{\natexlab{a}}),
  \eprint{0907.0462}.

\bibitem[{\citenamefont{{Pan} et~al.}(2010)\citenamefont{{Pan}, {Buonanno},
  {Buchman}, {Chu}, {Kidder}, {Pfeiffer}, and {Scheel}}}]{2010PhRvD..81h4041P}
\bibinfo{author}{\bibfnamefont{Y.}~\bibnamefont{{Pan}}},
  \bibinfo{author}{\bibfnamefont{A.}~\bibnamefont{{Buonanno}}},
  \bibinfo{author}{\bibfnamefont{L.~T.} \bibnamefont{{Buchman}}},
  \bibinfo{author}{\bibfnamefont{T.}~\bibnamefont{{Chu}}},
  \bibinfo{author}{\bibfnamefont{L.~E.} \bibnamefont{{Kidder}}},
  \bibinfo{author}{\bibfnamefont{H.~P.} \bibnamefont{{Pfeiffer}}},
  \bibnamefont{and} \bibinfo{author}{\bibfnamefont{M.~A.}
  \bibnamefont{{Scheel}}}, \bibinfo{journal}{\prd}
  \textbf{\bibinfo{volume}{81}}, \bibinfo{pages}{084041}
  (\bibinfo{year}{2010}), \eprint{0912.3466}.

\bibitem[{\citenamefont{{Flanagan} and {Hughes}}(1998)}]{1998PhRvD..57.4535F}
\bibinfo{author}{\bibfnamefont{{\'E}.~{\'E}.} \bibnamefont{{Flanagan}}}
  \bibnamefont{and} \bibinfo{author}{\bibfnamefont{S.~A.}
  \bibnamefont{{Hughes}}}, \bibinfo{journal}{\prd}
  \textbf{\bibinfo{volume}{57}}, \bibinfo{pages}{4535} (\bibinfo{year}{1998}).

\bibitem[{\citenamefont{{Boyle} and {Kesden}}(2008)}]{2008PhRvD..78b4017B}
\bibinfo{author}{\bibfnamefont{L.}~\bibnamefont{{Boyle}}} \bibnamefont{and}
  \bibinfo{author}{\bibfnamefont{M.}~\bibnamefont{{Kesden}}},
  \bibinfo{journal}{\prd} \textbf{\bibinfo{volume}{78}},
  \bibinfo{pages}{024017} (\bibinfo{year}{2008}).

\bibitem[{\citenamefont{{Boyle} et~al.}(2007)\citenamefont{{Boyle}, {Kesden},
  and {Nissanke}}}]{gr-nr-io-fitting-Boyle2007}
\bibinfo{author}{\bibfnamefont{L.}~\bibnamefont{{Boyle}}},
  \bibinfo{author}{\bibfnamefont{M.}~\bibnamefont{{Kesden}}}, \bibnamefont{and}
  \bibinfo{author}{\bibfnamefont{S.}~\bibnamefont{{Nissanke}}},
  \bibinfo{journal}{(arXiv:0709.0299)}  (\bibinfo{year}{2007}),
  \eprint{0709.0299}.

\bibitem[{\citenamefont{{Rezzolla}
  et~al.}(2008{\natexlab{a}})\citenamefont{{Rezzolla}, {Barausse}, {Dorband},
  {Pollney}, {Reisswig}, {Seiler}, and {Husa}}}]{2008PhRvD..78d4002R}
\bibinfo{author}{\bibfnamefont{L.}~\bibnamefont{{Rezzolla}}},
  \bibinfo{author}{\bibfnamefont{E.}~\bibnamefont{{Barausse}}},
  \bibinfo{author}{\bibfnamefont{E.~N.} \bibnamefont{{Dorband}}},
  \bibinfo{author}{\bibfnamefont{D.}~\bibnamefont{{Pollney}}},
  \bibinfo{author}{\bibfnamefont{C.}~\bibnamefont{{Reisswig}}},
  \bibinfo{author}{\bibfnamefont{J.}~\bibnamefont{{Seiler}}}, \bibnamefont{and}
  \bibinfo{author}{\bibfnamefont{S.}~\bibnamefont{{Husa}}},
  \bibinfo{journal}{\prd} \textbf{\bibinfo{volume}{78}},
  \bibinfo{pages}{044002} (\bibinfo{year}{2008}{\natexlab{a}}).

\bibitem[{\citenamefont{{Rezzolla}
  et~al.}(2008{\natexlab{b}})\citenamefont{{Rezzolla}, {Diener}, {Dorband},
  {Pollney}, {Reisswig}, {Schnetter}, and {Seiler}}}]{2008ApJ...674L..29R}
\bibinfo{author}{\bibfnamefont{L.}~\bibnamefont{{Rezzolla}}},
  \bibinfo{author}{\bibfnamefont{P.}~\bibnamefont{{Diener}}},
  \bibinfo{author}{\bibfnamefont{E.~N.} \bibnamefont{{Dorband}}},
  \bibinfo{author}{\bibfnamefont{D.}~\bibnamefont{{Pollney}}},
  \bibinfo{author}{\bibfnamefont{C.}~\bibnamefont{{Reisswig}}},
  \bibinfo{author}{\bibfnamefont{E.}~\bibnamefont{{Schnetter}}},
  \bibnamefont{and} \bibinfo{author}{\bibfnamefont{J.}~\bibnamefont{{Seiler}}},
  \bibinfo{journal}{\apjl} \textbf{\bibinfo{volume}{674}}, \bibinfo{pages}{L29}
  (\bibinfo{year}{2008}{\natexlab{b}}).

\bibitem[{\citenamefont{{Baker}
  et~al.}(2008{\natexlab{a}})\citenamefont{{Baker}, {Boggs}, {Centrella},
  {Kelly}, {McWilliams}, {Miller}, and {van
  Meter}}}]{gr-nr-io-fitting-GoddardGeneric2008}
\bibinfo{author}{\bibfnamefont{J.~G.} \bibnamefont{{Baker}}},
  \bibinfo{author}{\bibfnamefont{W.~D.} \bibnamefont{{Boggs}}},
  \bibinfo{author}{\bibfnamefont{J.}~\bibnamefont{{Centrella}}},
  \bibinfo{author}{\bibfnamefont{B.~J.} \bibnamefont{{Kelly}}},
  \bibinfo{author}{\bibfnamefont{S.~T.} \bibnamefont{{McWilliams}}},
  \bibinfo{author}{\bibfnamefont{M.~C.} \bibnamefont{{Miller}}},
  \bibnamefont{and} \bibinfo{author}{\bibfnamefont{J.~R.} \bibnamefont{{van
  Meter}}}, \bibinfo{journal}{ArXiv e-prints} \textbf{\bibinfo{volume}{802}}
  (\bibinfo{year}{2008}{\natexlab{a}}).

\bibitem[{\citenamefont{{Abadie et al.\ (The LIGO Scientific
  Collaboration)}}(2010)}]{LIGO-Inspiral-S5-Ranges}
\bibinfo{author}{\bibfnamefont{B.}~\bibnamefont{{Abadie et al.\ (The LIGO
  Scientific Collaboration)}}}, \bibinfo{journal}{(arXiv:1003.2481)}
  (\bibinfo{year}{2010}), \eprint{1003.2481}.

\bibitem[{\citenamefont{Herrmann et~al.}(2007)\citenamefont{Herrmann, Hinder,
  Shoemaker, Laguna, and Matzner}}]{Herrmann:2007ex}
\bibinfo{author}{\bibfnamefont{F.}~\bibnamefont{Herrmann}},
  \bibinfo{author}{\bibfnamefont{I.}~\bibnamefont{Hinder}},
  \bibinfo{author}{\bibfnamefont{D.~M.} \bibnamefont{Shoemaker}},
  \bibinfo{author}{\bibfnamefont{P.}~\bibnamefont{Laguna}}, \bibnamefont{and}
  \bibinfo{author}{\bibfnamefont{R.~A.} \bibnamefont{Matzner}},
  \bibinfo{journal}{Phys. Rev.} \textbf{\bibinfo{volume}{D76}},
  \bibinfo{pages}{084032} (\bibinfo{year}{2007}), \eprint{0706.2541}.

\bibitem[{\citenamefont{{Herrmann} et~al.}(2007)\citenamefont{{Herrmann},
  {Hinder}, {Shoemaker}, and {Laguna}}}]{2007CQGra..24...33H}
\bibinfo{author}{\bibfnamefont{F.}~\bibnamefont{{Herrmann}}},
  \bibinfo{author}{\bibfnamefont{I.}~\bibnamefont{{Hinder}}},
  \bibinfo{author}{\bibfnamefont{D.}~\bibnamefont{{Shoemaker}}},
  \bibnamefont{and} \bibinfo{author}{\bibfnamefont{P.}~\bibnamefont{{Laguna}}},
  \bibinfo{journal}{Classical and Quantum Gravity}
  \textbf{\bibinfo{volume}{24}}, \bibinfo{pages}{S33} (\bibinfo{year}{2007}),
  \eprint{arXiv:gr-qc/0601026}.

\bibitem[{\citenamefont{Herrmann et~al.}(2007)\citenamefont{Herrmann, Hinder,
  Shoemaker, Laguna, and Matzner}}]{Herrmann:2007ac}
\bibinfo{author}{\bibfnamefont{F.}~\bibnamefont{Herrmann}},
  \bibinfo{author}{\bibfnamefont{I.}~\bibnamefont{Hinder}},
  \bibinfo{author}{\bibfnamefont{D.}~\bibnamefont{Shoemaker}},
  \bibinfo{author}{\bibfnamefont{P.}~\bibnamefont{Laguna}}, \bibnamefont{and}
  \bibinfo{author}{\bibfnamefont{R.~A.} \bibnamefont{Matzner}},
  \bibinfo{journal}{Astrophys. J.} \textbf{\bibinfo{volume}{661}},
  \bibinfo{pages}{430} (\bibinfo{year}{2007}), \eprint{gr-qc/0701143}.

\bibitem[{\citenamefont{Hinder et~al.}(2008{\natexlab{a}})\citenamefont{Hinder,
  Vaishnav, Herrmann, Shoemaker, and Laguna}}]{Hinder:2007qu}
\bibinfo{author}{\bibfnamefont{I.}~\bibnamefont{Hinder}},
  \bibinfo{author}{\bibfnamefont{B.}~\bibnamefont{Vaishnav}},
  \bibinfo{author}{\bibfnamefont{F.}~\bibnamefont{Herrmann}},
  \bibinfo{author}{\bibfnamefont{D.}~\bibnamefont{Shoemaker}},
  \bibnamefont{and} \bibinfo{author}{\bibfnamefont{P.}~\bibnamefont{Laguna}},
  \bibinfo{journal}{Phys. Rev.} \textbf{\bibinfo{volume}{D77}},
  \bibinfo{pages}{081502} (\bibinfo{year}{2008}{\natexlab{a}}),
  \eprint{0710.5167}.

\bibitem[{\citenamefont{Healy et~al.}(2009{\natexlab{a}})}]{Healy:2008js}
\bibinfo{author}{\bibfnamefont{J.}~\bibnamefont{Healy}} \bibnamefont{et~al.},
  \bibinfo{journal}{Phys. Rev. Lett.} \textbf{\bibinfo{volume}{102}},
  \bibinfo{pages}{041101} (\bibinfo{year}{2009}{\natexlab{a}}),
  \eprint{0807.3292}.

\bibitem[{\citenamefont{Hinder et~al.}(2008{\natexlab{b}})\citenamefont{Hinder,
  Herrmann, Laguna, and Shoemaker}}]{Hinder:2008kv}
\bibinfo{author}{\bibfnamefont{I.}~\bibnamefont{Hinder}},
  \bibinfo{author}{\bibfnamefont{F.}~\bibnamefont{Herrmann}},
  \bibinfo{author}{\bibfnamefont{P.}~\bibnamefont{Laguna}}, \bibnamefont{and}
  \bibinfo{author}{\bibfnamefont{D.}~\bibnamefont{Shoemaker}},
  \textbf{\bibinfo{volume}{arxiv:0806.1037}}
  (\bibinfo{year}{2008}{\natexlab{b}}), \eprint{0806.1037}.

\bibitem[{\citenamefont{Healy et~al.}(2009{\natexlab{b}})\citenamefont{Healy,
  Levin, and Shoemaker}}]{Healy:2009zm}
\bibinfo{author}{\bibfnamefont{J.}~\bibnamefont{Healy}},
  \bibinfo{author}{\bibfnamefont{J.}~\bibnamefont{Levin}}, \bibnamefont{and}
  \bibinfo{author}{\bibfnamefont{D.}~\bibnamefont{Shoemaker}},
  \bibinfo{journal}{Phys. Rev. Lett.} \textbf{\bibinfo{volume}{103}},
  \bibinfo{pages}{131101} (\bibinfo{year}{2009}{\natexlab{b}}),
  \eprint{0907.0671}.

\bibitem[{\citenamefont{Healy et~al.}(2010)\citenamefont{Healy, Laguna,
  Matzner, and Shoemaker}}]{Healy:2009ir}
\bibinfo{author}{\bibfnamefont{J.}~\bibnamefont{Healy}},
  \bibinfo{author}{\bibfnamefont{P.}~\bibnamefont{Laguna}},
  \bibinfo{author}{\bibfnamefont{R.~A.} \bibnamefont{Matzner}},
  \bibnamefont{and} \bibinfo{author}{\bibfnamefont{D.~M.}
  \bibnamefont{Shoemaker}}, \bibinfo{journal}{Phys. Rev.}
  \textbf{\bibinfo{volume}{D81}}, \bibinfo{pages}{081501}
  (\bibinfo{year}{2010}), \eprint{0905.3914}.

\bibitem[{\citenamefont{Bode et~al.}(2010)\citenamefont{Bode, Haas, Bogdanovic,
  Laguna, and Shoemaker}}]{Bode:2009mt}
\bibinfo{author}{\bibfnamefont{T.}~\bibnamefont{Bode}},
  \bibinfo{author}{\bibfnamefont{R.}~\bibnamefont{Haas}},
  \bibinfo{author}{\bibfnamefont{T.}~\bibnamefont{Bogdanovic}},
  \bibinfo{author}{\bibfnamefont{P.}~\bibnamefont{Laguna}}, \bibnamefont{and}
  \bibinfo{author}{\bibfnamefont{D.}~\bibnamefont{Shoemaker}},
  \bibinfo{journal}{Astrophys. J.} \textbf{\bibinfo{volume}{715}},
  \bibinfo{pages}{1117} (\bibinfo{year}{2010}), \eprint{0912.0087}.

\bibitem[{\citenamefont{Schnetter et~al.}(2004)\citenamefont{Schnetter, Hawley,
  and Hawke}}]{Schnetter-etal-03b}
\bibinfo{author}{\bibfnamefont{E.}~\bibnamefont{Schnetter}},
  \bibinfo{author}{\bibfnamefont{S.~H.} \bibnamefont{Hawley}},
  \bibnamefont{and} \bibinfo{author}{\bibfnamefont{I.}~\bibnamefont{Hawke}},
  \bibinfo{journal}{Class. Quant. Grav.} \textbf{\bibinfo{volume}{21}},
  \bibinfo{pages}{1465} (\bibinfo{year}{2004}).

\bibitem[{cactus-web()}]{cactus-web}
cactus-web, \bibinfo{note}{cactus Computational Toolkit home page:\\{\tt
  http://www.cactuscode.org}}.

\bibitem[{\citenamefont{Husa et~al.}(2006)\citenamefont{Husa, Hinder, and
  Lechner}}]{Husa:2004ip}
\bibinfo{author}{\bibfnamefont{S.}~\bibnamefont{Husa}},
  \bibinfo{author}{\bibfnamefont{I.}~\bibnamefont{Hinder}}, \bibnamefont{and}
  \bibinfo{author}{\bibfnamefont{C.}~\bibnamefont{Lechner}},
  \bibinfo{journal}{Computer Physics Communications}
  \textbf{\bibinfo{volume}{174}}, \bibinfo{pages}{983} (\bibinfo{year}{2006}).

\bibitem[{\citenamefont{{Baker} et~al.}(2007)\citenamefont{{Baker},
  {McWilliams}, {Meter}, {Centrella}, {Choi}, {Kelly}, and
  {Koppitz}}}]{nr-Goddard-EarlyInspiralSummary-2007}
\bibinfo{author}{\bibfnamefont{J.~G.} \bibnamefont{{Baker}}},
  \bibinfo{author}{\bibfnamefont{S.~T.} \bibnamefont{{McWilliams}}},
  \bibinfo{author}{\bibfnamefont{J.~R.~v.} \bibnamefont{{Meter}}},
  \bibinfo{author}{\bibfnamefont{J.}~\bibnamefont{{Centrella}}},
  \bibinfo{author}{\bibfnamefont{D.}~\bibnamefont{{Choi}}},
  \bibinfo{author}{\bibfnamefont{B.~J.} \bibnamefont{{Kelly}}},
  \bibnamefont{and}
  \bibinfo{author}{\bibfnamefont{M.}~\bibnamefont{{Koppitz}}},
  \bibinfo{journal}{\prd} \textbf{\bibinfo{volume}{75}},
  \bibinfo{pages}{124024} (\bibinfo{year}{2007}),
  \urlprefix\url{http://xxx.lanl.gov/abs/gr-qc/0612117}.

\bibitem[{\citenamefont{{Scheel} et~al.}(2008)\citenamefont{{Scheel}, {Boyle},
  {Chu}, {Kidder}, {Matthews}, and {Pfeiffer}}}]{gwastro-cornell-highres2008}
\bibinfo{author}{\bibfnamefont{M.~A.} \bibnamefont{{Scheel}}},
  \bibinfo{author}{\bibfnamefont{M.}~\bibnamefont{{Boyle}}},
  \bibinfo{author}{\bibfnamefont{T.}~\bibnamefont{{Chu}}},
  \bibinfo{author}{\bibfnamefont{L.~E.} \bibnamefont{{Kidder}}},
  \bibinfo{author}{\bibfnamefont{K.~D.} \bibnamefont{{Matthews}}},
  \bibnamefont{and} \bibinfo{author}{\bibfnamefont{H.~P.}
  \bibnamefont{{Pfeiffer}}}, \bibinfo{journal}{ArXiv e-prints}
  (\bibinfo{year}{2008}).

\bibitem[{\citenamefont{{Baker}
  et~al.}(2008{\natexlab{b}})\citenamefont{{Baker}, {Boggs}, {Centrella},
  {Kelly}, {McWilliams}, and {van Meter}}}]{2008PhRvD..78d4046B}
\bibinfo{author}{\bibfnamefont{J.~G.} \bibnamefont{{Baker}}},
  \bibinfo{author}{\bibfnamefont{W.~D.} \bibnamefont{{Boggs}}},
  \bibinfo{author}{\bibfnamefont{J.}~\bibnamefont{{Centrella}}},
  \bibinfo{author}{\bibfnamefont{B.~J.} \bibnamefont{{Kelly}}},
  \bibinfo{author}{\bibfnamefont{S.~T.} \bibnamefont{{McWilliams}}},
  \bibnamefont{and} \bibinfo{author}{\bibfnamefont{J.~R.} \bibnamefont{{van
  Meter}}}, \bibinfo{journal}{\prd} \textbf{\bibinfo{volume}{78}},
  \bibinfo{pages}{044046} (\bibinfo{year}{2008}{\natexlab{b}}).

\bibitem[{\citenamefont{{Vaishnav} et~al.}(2007)\citenamefont{{Vaishnav},
  {Hinder}, {Herrmann}, and {Shoemaker}}}]{2007PhRvD..76h4020V}
\bibinfo{author}{\bibfnamefont{B.}~\bibnamefont{{Vaishnav}}},
  \bibinfo{author}{\bibfnamefont{I.}~\bibnamefont{{Hinder}}},
  \bibinfo{author}{\bibfnamefont{F.}~\bibnamefont{{Herrmann}}},
  \bibnamefont{and}
  \bibinfo{author}{\bibfnamefont{D.}~\bibnamefont{{Shoemaker}}},
  \bibinfo{journal}{\prd} \textbf{\bibinfo{volume}{76}},
  \bibinfo{pages}{084020} (\bibinfo{year}{2007}).

\bibitem[{\citenamefont{{Shoemaker} et~al.}(2008)\citenamefont{{Shoemaker},
  {Vaishnav}, {Hinder}, and {Herrmann}}}]{2008CQGra..25k4047S}
\bibinfo{author}{\bibfnamefont{D.}~\bibnamefont{{Shoemaker}}},
  \bibinfo{author}{\bibfnamefont{B.}~\bibnamefont{{Vaishnav}}},
  \bibinfo{author}{\bibfnamefont{I.}~\bibnamefont{{Hinder}}}, \bibnamefont{and}
  \bibinfo{author}{\bibfnamefont{F.}~\bibnamefont{{Herrmann}}},
  \bibinfo{journal}{Classical and Quantum Gravity}
  \textbf{\bibinfo{volume}{25}}, \bibinfo{pages}{114047}
  (\bibinfo{year}{2008}).

\bibitem[{\citenamefont{{Ajith} et~al.}(2007)\citenamefont{{Ajith}, {Babak},
  {Chen}, {Hewitson}, {Krishnan}, {Whelan}, {Br{\"u}gmann}, {Diener},
  {Gonzalez}, {Hannam} et~al.}}]{nr-Jena-nonspinning-templates2007}
\bibinfo{author}{\bibfnamefont{P.}~\bibnamefont{{Ajith}}},
  \bibinfo{author}{\bibfnamefont{S.}~\bibnamefont{{Babak}}},
  \bibinfo{author}{\bibfnamefont{Y.}~\bibnamefont{{Chen}}},
  \bibinfo{author}{\bibfnamefont{M.}~\bibnamefont{{Hewitson}}},
  \bibinfo{author}{\bibfnamefont{B.}~\bibnamefont{{Krishnan}}},
  \bibinfo{author}{\bibfnamefont{J.~T.} \bibnamefont{{Whelan}}},
  \bibinfo{author}{\bibfnamefont{B.}~\bibnamefont{{Br{\"u}gmann}}},
  \bibinfo{author}{\bibfnamefont{P.}~\bibnamefont{{Diener}}},
  \bibinfo{author}{\bibfnamefont{J.}~\bibnamefont{{Gonzalez}}},
  \bibinfo{author}{\bibfnamefont{M.}~\bibnamefont{{Hannam}}},
  \bibnamefont{et~al.}, \bibinfo{journal}{Classical and Quantum Gravity}
  \textbf{\bibinfo{volume}{24}}, \bibinfo{pages}{689} (\bibinfo{year}{2007}),
  \eprint{0704.3764}.

\bibitem[{\citenamefont{{Ajith} et~al.}(2008)\citenamefont{{Ajith}, {Babak},
  {Chen}, {Hewitson}, {Krishnan}, {Sintes}, {Whelan}, {Br{\"u}gmann}, {Diener},
  {Dorband} et~al.}}]{2008PhRvD..77j4017A}
\bibinfo{author}{\bibfnamefont{P.}~\bibnamefont{{Ajith}}},
  \bibinfo{author}{\bibfnamefont{S.}~\bibnamefont{{Babak}}},
  \bibinfo{author}{\bibfnamefont{Y.}~\bibnamefont{{Chen}}},
  \bibinfo{author}{\bibfnamefont{M.}~\bibnamefont{{Hewitson}}},
  \bibinfo{author}{\bibfnamefont{B.}~\bibnamefont{{Krishnan}}},
  \bibinfo{author}{\bibfnamefont{A.~M.} \bibnamefont{{Sintes}}},
  \bibinfo{author}{\bibfnamefont{J.~T.} \bibnamefont{{Whelan}}},
  \bibinfo{author}{\bibfnamefont{B.}~\bibnamefont{{Br{\"u}gmann}}},
  \bibinfo{author}{\bibfnamefont{P.}~\bibnamefont{{Diener}}},
  \bibinfo{author}{\bibfnamefont{N.}~\bibnamefont{{Dorband}}},
  \bibnamefont{et~al.}, \bibinfo{journal}{\prd} \textbf{\bibinfo{volume}{77}},
  \bibinfo{pages}{104017} (\bibinfo{year}{2008}).

\bibitem[{\citenamefont{{Sturani} et~al.}(2010)\citenamefont{{Sturani},
  {Fischetti}, {Cadonati}, {Guidi}, {Healy}, {Shoemaker}, and
  {Vicer{\'e}}}}]{2010arXiv1005.0551S}
\bibinfo{author}{\bibfnamefont{R.}~\bibnamefont{{Sturani}}},
  \bibinfo{author}{\bibfnamefont{S.}~\bibnamefont{{Fischetti}}},
  \bibinfo{author}{\bibfnamefont{L.}~\bibnamefont{{Cadonati}}},
  \bibinfo{author}{\bibfnamefont{G.~M.} \bibnamefont{{Guidi}}},
  \bibinfo{author}{\bibfnamefont{J.}~\bibnamefont{{Healy}}},
  \bibinfo{author}{\bibfnamefont{D.}~\bibnamefont{{Shoemaker}}},
  \bibnamefont{and}
  \bibinfo{author}{\bibfnamefont{A.}~\bibnamefont{{Vicer{\'e}}}},
  \bibinfo{journal}{ArXiv e-prints}  (\bibinfo{year}{2010}),
  \eprint{1005.0551}.

\bibitem[{\citenamefont{{Reisswig}
  et~al.}(2009{\natexlab{b}})\citenamefont{{Reisswig}, {Bishop}, {Pollney}, and
  {Szilagyi}}}]{gr-nr-methods-ExtractionAtInfinity-Reisswig2009}
\bibinfo{author}{\bibfnamefont{C.}~\bibnamefont{{Reisswig}}},
  \bibinfo{author}{\bibfnamefont{N.~T.} \bibnamefont{{Bishop}}},
  \bibinfo{author}{\bibfnamefont{D.}~\bibnamefont{{Pollney}}},
  \bibnamefont{and}
  \bibinfo{author}{\bibfnamefont{B.}~\bibnamefont{{Szilagyi}}},
  \bibinfo{journal}{(arXiv:0912.1285)}  (\bibinfo{year}{2009}{\natexlab{b}}),
  \urlprefix\url{http://xxx.lanl.gov/abs/arXiv:0912.1285}.

\bibitem[{\citenamefont{{Amaro-Seoane} and
  {Santamaria}}(2009)}]{gwastro-imbh-ComparableMergers-2009}
\bibinfo{author}{\bibfnamefont{P.}~\bibnamefont{{Amaro-Seoane}}}
  \bibnamefont{and}
  \bibinfo{author}{\bibfnamefont{L.}~\bibnamefont{{Santamaria}}},
  \bibinfo{journal}{ArXiv e-prints}  (\bibinfo{year}{2009}).

\bibitem[{\citenamefont{{Mandel} et~al.}(2009)\citenamefont{{Mandel}, {Gair},
  and {Miller}}}]{gwastro-imbh-ComparableMergers-Ilya-2009}
\bibinfo{author}{\bibfnamefont{I.}~\bibnamefont{{Mandel}}},
  \bibinfo{author}{\bibfnamefont{J.~R.} \bibnamefont{{Gair}}},
  \bibnamefont{and} \bibinfo{author}{\bibfnamefont{M.~C.}
  \bibnamefont{{Miller}}}, \bibinfo{journal}{ArXiv e-prints}
  (\bibinfo{year}{2009}).

\bibitem[{\citenamefont{{Gair} et~al.}(2009)\citenamefont{{Gair}, {Mandel},
  {Miller}, and {Volonteri}}}]{gw-astro-ET-IMBH-ReviewMiller-2009}
\bibinfo{author}{\bibfnamefont{J.~R.} \bibnamefont{{Gair}}},
  \bibinfo{author}{\bibfnamefont{I.}~\bibnamefont{{Mandel}}},
  \bibinfo{author}{\bibfnamefont{M.~C.} \bibnamefont{{Miller}}},
  \bibnamefont{and}
  \bibinfo{author}{\bibfnamefont{M.}~\bibnamefont{{Volonteri}}},
  \bibinfo{journal}{(arXiv:0907.5450)}  (\bibinfo{year}{2009}),
  \urlprefix\url{http://xxx.lanl.gov/abs/arXiv:0907.5450}.

\bibitem[{\citenamefont{{Harry} and {the LIGO Scientific
  Collaboration}}(2010)}]{2010CQGra..27h4006H}
\bibinfo{author}{\bibfnamefont{G.~M.} \bibnamefont{{Harry}}} \bibnamefont{and}
  \bibinfo{author}{\bibnamefont{{the LIGO Scientific Collaboration}}},
  \bibinfo{journal}{Classical and Quantum Gravity}
  \textbf{\bibinfo{volume}{27}}, \bibinfo{pages}{084006}
  (\bibinfo{year}{2010}).

\bibitem[{\citenamefont{{Sathyaprakash} and {et.
  al}}(2009)}]{gw-detectors-ET-ScienceDocument}
\bibinfo{author}{\bibfnamefont{B.}~\bibnamefont{{Sathyaprakash}}}
  \bibnamefont{and} \bibinfo{author}{\bibnamefont{{et. al}}}
  (\bibinfo{year}{2009}),
  \urlprefix\url{https://workarea.et-gw.eu/et/WG4-Astrophysics/visdoc/}.

\bibitem[{\citenamefont{{Cutler} and {Flanagan}}(1994)}]{CutlerFlanagan:1994}
\bibinfo{author}{\bibfnamefont{C.}~\bibnamefont{{Cutler}}} \bibnamefont{and}
  \bibinfo{author}{\bibfnamefont{E.}~\bibnamefont{{Flanagan}}},
  \bibinfo{journal}{\prd} \textbf{\bibinfo{volume}{49}}, \bibinfo{pages}{2658}
  (\bibinfo{year}{1994}), \eprint{gr-qc/9402014}.

\end{thebibliography}
